\documentclass[11pt,reqno,preprint,dvipsnames]{article}
\usepackage{jheppub,epsfig,amssymb,amsmath,mathrsfs,hyperref,array,hhline}
\usepackage[makeroom]{cancel}
\usepackage{multirow,bbm}
\usepackage{feynmp-auto}
\usepackage{cleveref}
\usepackage[textsize=tiny]{todonotes}
\usepackage{mathtools}
\usepackage{caption}
\usepackage{subcaption}
\usepackage{placeins}
\usepackage{marginnote}
\makeatletter
\DeclareRobustCommand*{\bfseries}{%
  \not@math@alphabet\bfseries\mathbf
  \fontseries\bfdefault\selectfont
  \boldmath
}

\def\e{\epsilon}
\def\be{\begin{equation}}
\def\ee{\end{equation}}
\newcommand{\bea}{\begin{eqnarray}}
\newcommand{\eea}{\end{eqnarray}}

\def\eqn#1{eq.~\eqref{#1}}

\def\eqns#1#2{eqs.~(\ref{#1}) and~(\ref{#2})}

\def\blue#1{{\color{blue}#1}}
\def\green#1{{\color{green}#1}}

\def\disc{\text{Disc}}
\def\Gcusp{\Gamma_{\rm cusp}}

\def\Li{\textrm{Li}}

\def\EE{{\cal E}}

\newcommand{\cC}{\mathcal{C}}

\newcommand{\cL}{\begin{cal}L\end{cal}}

\def\blue#1{{\color{blue}#1}}
\def\red#1{{\color{red}#1}}

\title{Bootstrapping a Stress-Tensor Form Factor through Eight Loops}

\author{Lance~J.~Dixon$^{1}$, {\"O}mer G{\"u}rdo{\u{g}}an$^{2}$, Andrew~J.~McLeod$^{3,4,5}$ and Matthias Wilhelm$^{5}$}

\affiliation{$^1$ SLAC National Accelerator Laboratory,
Stanford University, Stanford, CA 94309, USA}

\affiliation{$^2$ School of Physics \& Astronomy, 
University of Southampton, Southampton, SO17 1BJ, UK}

\affiliation{$^3$ CERN, Theoretical Physics Department, 1211 Geneva 23, Switzerland}

\affiliation{$^4$ Mani L. Bhaumik Institute for Theoretical Physics, \\ UCLA Department of Physics and Astronomy, Los Angeles, CA 90095, USA}

\affiliation{$^5$ Niels Bohr International Academy, Niels Bohr Institute,
  Blegdamsvej 17, 2100 Copenhagen \O{}, Denmark}

\abstract{We bootstrap the three-point form factor of the chiral stress-tensor multiplet in planar $\mathcal{N}=4$ supersymmetric Yang-Mills theory at six, seven, and eight loops, using boundary data from the form factor operator product expansion.  This may represent the highest perturbative order to which multi-variate quantities in a unitary four-dimensional quantum field theory have been computed. In computing this form factor, we observe and employ new restrictions on pairs and triples of adjacent letters in the symbol.  We provide details about the function space required to describe the form factor through eight loops.  Plotting the results on various lines provides striking numerical evidence for a finite radius of convergence of perturbation theory.
By the principle of maximal transcendentality, our results are expected to give the highest weight part of the $g g \to H g$ and $H \to ggg$ amplitudes in the heavy-top limit of QCD through eight loops. These results were also recently used to discover a new antipodal duality between this form factor and a six-point amplitude in the same theory. }

\preprint{ \begin{flushright} SLAC-PUB-17653 \\ CERN-TH-2022-039 \end{flushright}}

\begin{document}
\maketitle
\flushbottom
\begin{fmffile}{feyndiags}


\section{Introduction}
\label{sec:introduction}

Significant progress has been made in recent decades in the development of general methods for carrying out perturbative calculations in quantum field theory. Even so, these general methods remain limited to low loop orders, as they rely on evaluating Feynman integrals, which quickly become prohibitively difficult to compute using current technology. 

As a result, in cases where the analytic properties of a quantity are (conjecturally) understood to high loop order, more progress can sometimes be made by adopting a bootstrap method that bypasses integration altogether. In this approach, one constructs the space of functions within which a quantity is believed to lie, and attempts to identify a unique function in this space that has all of its expected properties. Such an approach has proven highly successful for determining amplitudes in planar $\mathcal{N}=4$ supersymmetric Yang-Mills (sYM) theory~\cite{Dixon:2011pw,Dixon:2011nj,Dixon:2013eka,Dixon:2014voa,Dixon:2014iba,Drummond:2014ffa,Dixon:2015iva,Caron-Huot:2016owq,Dixon:2016apl,Dixon:2016nkn,Drummond:2018caf,Caron-Huot:2019vjl}, as well as soft anomalous dimensions~\cite{Li:2016ctv,Almelid:2017qju}.

The three-point form factor of the chiral part of the stress-tensor supermultiplet in planar $\mathcal{N}=4$ sYM theory is another quantity that turns out to be amenable to bootstrap techniques. This was first shown at two loops~\cite{Brandhuber:2012vm}, and more recently input from the form-factor operator product expansion (FFOPE)~\cite{Sever:2020jjx,Sever:2021nsq,Sever:2021xga} was used to bootstrap this form factor through five loops~\cite{Dixon:2020bbt}.\footnote{The results of our bootstrap procedure have been cross-checked numerically at three loops~\cite{Lin:2021qol}.}
In this paper, we make use of a further-improved bootstrap approach to calculate the three-point form factor at six, seven, and eight loops.

Our bootstrap method takes as its starting point several observations about the mathematical structure of the two-loop three-point form factor, which we assume generalize to higher loop orders. 
In particular, our starting assumption is that the space of two-dimensional harmonic polylogarithms (2dHPLs)~\cite{Gehrmann:2000zt} remains sufficient for expressing the three-point form factor perturbatively to all loop orders. These functions form a subspace of the multiple polylogarithms, and as such come equipped with a Hopf algebra structure \cite{Gonch2,Brown:2011ik,Duhr:2011zq,Duhr:2012fh}, which includes the symbol map~\cite{Goncharov:2010jf}.
We also assume the extended Steinmann-like (ES-like) conditions that were previously observed on adjacent symbol entries~\cite{Dixon:2020bbt,Chicherin:2020umh} remain true at higher loop orders,
and here identify \emph{new} ES-like conditions that we additionally impose. Furthermore, we observe new restrictions on the adjacent triples of letters that appear in the symbol,
as well as multiple-final-entry conditions that extend those previously found in ref.~\cite{Dixon:2020bbt}.  
Finally, we restrict the functions that can appear in the first entry of the coproduct of these form factors (when they are considered in an appropriate normalization) in accordance with a coaction principle, similar to what was done in bootstrapping the six-point amplitude of planar $\mathcal{N}=4$ sYM theory~\cite{Caron-Huot:2019bsq}.

The results obtained in this paper have been instrumental in discovering and providing evidence for a surprising new duality \cite{Dixon:2021tdw} between the three-point maximally helicity violating (MHV) form factor under consideration and the six-point MHV scattering amplitude in the same theory, which has previously been bootstrapped up to seven-loop order \cite{Caron-Huot:2019vjl}. Concretely, the (cosmically normalized) three-point form factor, which is a function of two variables, is mapped to the six-point amplitude evaluated on a two-dimensional parity-even surface via the antipode map defined on polylogarithms~\cite{Dixon:2021tdw}. At symbol level, the antipode simply acts by reversing the order of the symbol entries (up to an overall sign that depends on the transcendental weight)~\cite{Gonch3,Brown:2013gia}.
This means that the antipodal duality relates the discontinuities of the form factor, which are encoded in its first symbol entry, to the derivative of the amplitude, which are encoded in its last symbol entry, and vice versa.
By means of the duality between amplitudes and form factors to Wilson loops~\cite{Alday:2007hr,Drummond:2007aua,Brandhuber:2007yx,Drummond:2007cf,Drummond:2007au,Alday:2008yw,Adamo:2011pv,Alday:2007he,Maldacena:2010kp,Brandhuber:2010ad}, the new antipodal duality can also be viewed as a map between periodic Wilson loops formed by three unique light-like edges, and specific configurations of closed hexagonal light-like Wilson loops. Currently, we have no physical understanding of (or derivation for) why the antipodal duality should hold, despite the  overwhelming empirical evidence for it up to seven loops~\cite{Dixon:2021tdw}.

We remark that this antipodal duality was discovered only after the form factor was completely determined via the bootstrap computation that we report on in this paper through eight loops. Even so, this duality provides a useful perspective on the empirical adjacency and multiple-final-entry conditions that we have observed.
In particular, when combined with the antipodal duality, the ES-like and triplet adjacency conditions on the form factor follow from the extended Steinmann relations and the integrability conditions of the six-point amplitude.
Similarly, the multiple-final-entry conditions observed for the form factor follow from the combined power of the six-particle branch cut conditions, the extended Steinmann conditions, and (starting at weight eight) the implications of the pentagon operator product expansion (POPE)~\cite{Alday:2010ku,Basso:2013vsa,Basso:2013aha,Basso:2014koa,Basso:2014jfa,Basso:2014nra,Belitsky:2014sla,Belitsky:2014lta,Basso:2014hfa,Belitsky:2015efa,Basso:2015rta,Basso:2015uxa,Belitsky:2016vyq} for the six-point amplitude, which restricts what transcendental constants can appear at specific points in the space of six-point kinematics~\cite{Caron-Huot:2019bsq}.

The remainder of this paper is structured as follows.
In section \ref{sec:background}, we review some facts about the three-point form factor.
We then describe our bootstrap procedure in section \ref{sec: bootstrap procedure}.
In section \ref{sec:results}, we describe some of the constraints in more detail and present our results for the form factor at six, seven and eight loops.
We elaborate on the aforementioned consequences of the antipodal duality in section \ref{sec:duality}.
In section \ref{sec:special_kinematics}, we study our results at interesting lines and points in the kinematic space.
We conclude with a summary and outlook on further interesting questions in section \ref{sec:conclusions}.

The main results of the paper are included as ancillary files, only some of which have been small enough to be included on the arXiv. The files that we have packaged with the arXiv submission include {\tt Esymb.txt}, which gives the symbol of the function $\mathcal{E}^{(L)}$ through seven loops, a file {\tt FFmultifinalentry.txt} that describes the space of multiple-final-entries that $\mathcal{E}^{(L)}$ satisfies through eight loops, and a pair of files {\tt T2termsyser8loops.txt} and {\tt T4termsyser8loops.txt} that give the near-collinear expansion of the form factor as a series expansion in the variable $y = S^2$ at order $T^2$ and $T^4$. Additional results have also been made available at~\cite{erda}. At this website, one can find the $T^2$ and $T^4$ terms before series expansion in the files {\tt T2terms8loops.txt} and {\tt T4terms8loops.txt}.  We also provide the symbol of the eight-loop form factor in the file {\tt Esymboct8.txt}, as well as the full function-level result for the form factor through eight loops in {\tt E\_C.txt}. The latter file refers to three additional files {\tt Cfns.txt}, {\tt Ccop8.txt}, and {\tt E100.txt} (all available at the same website) for the iterative definition of the $\mathcal{C}$ function space through weight eight, and the boundary values of the independent coproduct entries that appear in the form factor.


\section{The Three-Point Form Factor and Polylogarithms}
\label{sec:background}

In this paper, we study the three-point form factor defined by an insertion of the chiral part of the stress-tensor supermultiplet in planar $\mathcal{N} = 4$ sYM theory. 
The only helicity configurations this form factor can take are MHV and $\overline{\text{MHV}}$, which are related by a combination of parity, supersymmetry, and R-symmetry.
We thus restrict our attention to the MHV configuration, which we denote by $\mathcal{F}_3$.  The formal definition of $\mathcal{F}_3$ can be given in harmonic superspace, as for instance found in refs.~\cite{Eden:2011yp,Brandhuber:2011tv,Bork:2014eqa}.

The kinematic dependence of $\mathcal{F}_3$ can be expressed entirely in terms of the Mandelstam invariants $s_{12}$, $s_{23}$, $s_{31}$, and $s_{123}$, defined by
\begin{equation}
s_{i\dots j} = (p_i + p_{i+1} + \cdots + p_j)^2 \, ,
\end{equation} 
where $p_1$, $p_2$, and $p_3$ are the momenta associated with the three on-shell external states.
By momentum conservation, $s_{123}$ is equal to the invariant mass associated with the operator insertion. Since the on-shell external states are all massless, there also exists a linear relation between these invariants, namely
\begin{equation} \label{eq:mandelstam_relation}
q^2 = s_{123}  = s_{12} + s_{23} + s_{31} \, ,
\end{equation}
where $q$ denotes the momentum of the operator insertion.

An overall rational prefactor for $\mathcal{F}_3$ can be factored out of its weak-coupling expansion, leaving just a pure transcendental contribution at each loop order. This transcendental contribution only depends on dimensionless arguments, motivating us to define the set of rescaled variables
\begin{equation} \label{eq:uvwdef}
u = \frac{s_{12}}{q^2}, \qquad v = \frac{s_{23}}{q^2}, \qquad w = \frac{s_{31}}{q^2}\,, \qquad u+ v+w=1 . 
\end{equation}  
Only two of these variables are independent due to the relation~\eqref{eq:mandelstam_relation}, but we will often make use of all three since this makes the dihedral symmetry of $\mathcal{F}_3$ manifest.

In addition to separating out the rational prefactor, we can define an infrared-finite version of $\mathcal{F}_3$ by dividing out an infrared-divergent factor $\mathcal{F}_3^{\text{BDS}}$ analogous to the BDS ansatz for amplitudes~\cite{Bern:2005iz,Brandhuber:2012vm}. The function $\mathcal{F}_3^{\text{BDS}}$ is given by the exponential of the one-loop contribution to $\mathcal{F}_3$~\cite{Brandhuber:2010ad,Brandhuber:2012vm}, namely
\begin{align}
M_3^{(1)}(\e,s_{ij})
&= - \frac{1}{\e^2} \sum_{i=1}^3 \left(\frac{\mu^2}{-s_{i,i+1}}\right)^\e 
- 2 \big( \Li_2(1-u) + \Li_2(1-v) + \Li_2(1-w) \big) \\
&\qquad \qquad - \ln u \ln v - \ln v \ln w - \ln w\ln u + \frac{9}{2} \zeta_2  \, , \nonumber
\end{align}
dressed by certain kinematic-independent anomalous dimensions. The finite function that remains is encoded by the remainder function $R_3$, defined via the relation 
\begin{equation}
\mathcal{F}_3 = \mathcal{F}_3^{\text{BDS}} \exp \left[ R_3 \right] \, .
\end{equation}
Expanding $R \equiv R_3$ in the 't Hooft coupling $\smash{g^2 = \frac{g^2_{\text{YM}} N_c}{16 \pi^2}}$, where the gauge group is SU($N_c$), with $N_c\to\infty$ at fixed $g^2$, we define the $L$-loop contribution $R^{(L)}$ to be the coefficient of $g^{2L}$, namely
\begin{equation}
R = \sum_{L=2}^\infty g^{2L} R^{(L)} \, .
\end{equation}
The coefficients $R^{(L)}$ can be considered to be functions of any two of the kinematic variables in eq.~\eqref{eq:uvwdef}, since the third variable can be eliminated using the relation~\eqref{eq:mandelstam_relation}.

The two-loop remainder function $R^{(2)}$ was computed in ref.~\cite{Brandhuber:2012vm}, where it was observed to be expressible in terms of classical polylogarithms. More generally, based on the observed properties of high-loop amplitudes in this theory~\cite{Dixon:2013eka,Dixon:2014iba,Dixon:2014voa,Drummond:2014ffa,Dixon:2015iva,Caron-Huot:2016owq,Dixon:2016nkn,Drummond:2018caf,Caron-Huot:2019vjl,Dixon:2020cnr}, $R^{(L)}$ is expected to be expressible in terms of multiple polylogarithms of uniform transcendental weight $2L$. Functions of this type can be defined in terms of a choice of integration base point and their (iterated) total differential, which takes the form 
\begin{equation} \label{eq:MPL_differential}
d F = \sum_{\phi\in\cL} F^\phi\, d \ln \phi \, ,
\end{equation} 
where the functions $F^\phi$ are multiple polylogarithms whose transcendental weight is one lower than that of $F$, and the sum is over some independent set of algebraic functions $\phi$. The transcendental weight of the natural logarithm is one, while the weight of an algebraic function is zero.

Most of the information about a multiple polylogarithm $F$ is encoded in its symbol $\mathcal{S}(F)$, which is defined by promoting the $d\ln$ factors in eq.~\eqref{eq:MPL_differential} to elements in a tensor product via the recursive definition~\cite{Goncharov:2010jf}
\begin{equation} \label{eq:symbol}
\mathcal{S}(F) = \sum_{\phi\in\cL} \mathcal{S}(F^\phi\,) \otimes \phi \, .
\end{equation} 
The algebraic functions $\phi$ that appear in the symbol are referred to as symbol letters, and their union $\cL$ is referred to as the symbol alphabet of $F$. The symbol alphabet of $R^{(2)}$ can be put in the form
\begin{equation} \label{eq:uv_letters}
\cL_{u} = \{u,v,w,1{-}u,1{-}v,1{-}w  \} = \{u,v,1{-}u{-}v,1{-}u,1{-}v,u{+}v  \}  \, .
\end{equation}
The first form of these letters exposes the dihedral symmetry of the alphabet, while the second shows that they can be expressed in terms of the more restrictive class of 2dHPLs~\cite{Gehrmann:2000zt}.

More recently, starting from a polylogarithmic ansatz built using the assumption that no new symbol letters appear in the remainder function at higher loops, the three-point form factor was bootstrapped through five loops~\cite{Dixon:2020bbt}. In that work, it was observed that a different choice of infrared normalization could be used to define a function other than the remainder function that exhibits a number of advantageous analytic properties. This new normalization corresponds to normalizing by~\cite{Dixon:2020bbt} a BDS-like ansatz analogous to the one for amplitudes~\cite{Alday:2009dv,Yang:2010as,Dixon:2015iva,Caron-Huot:2016owq,Dixon:2016nkn}:
\begin{equation}
\mathcal{F}_3 = \mathcal{F}_3^{\text{BDS-like}} \times \mathcal{E } \,.
\end{equation}
Here, $\mathcal{E}$ is related to $R$ by
\begin{equation} \label{eq:calE_def}
\mathcal{E} = \exp \left[ \frac{1}{4} \Gcusp \mathcal{E}^{(1)} + R\right] \, ,
\end{equation} 
where 
\begin{equation}
\mathcal{E}^{(1)}(u,v,w) = 2 \Li_2 \left(1 - \frac 1 u \right) + 2 \Li_2 \left(1 - \frac 1 v \right) + 2 \Li_2 \left(1 - \frac 1 w \right) \, ,
\label{eq:E_1loop}
\end{equation}
and 
\begin{equation}
\Gcusp  = 4 g^2 - 8 \zeta_2 g^4 + 88 \zeta_4 g^6 - 4\left[ 219 \zeta_6 + 8 (\zeta_3)^2 \right] g^8 + \dots
\end{equation}
is the cusp anomalous dimension~\cite{Beisert:2006ez}. Like the remainder function, $\mathcal{E}$ can be expanded at weak coupling as
\begin{equation}
\mathcal{E} = \sum_{L=1}^\infty g^{2L} \mathcal{E}^{(L)} \, 
\end{equation}
to define the $L$-loop contribution $\mathcal{E}^{(L)}$.

The primary benefit of working with the function $\mathcal{E}^{(L)}$ rather than $R^{(L)}$ is that the former respects certain non-adjacency relations in the symbol. To describe these constraints, it proves advantageous to define a different symbol alphabet from the one chosen above, which we denote by
\begin{equation} \label{eq:abc_letters}
\cL_{a} = \{a,b,c,d,e,f \} \, ,
\end{equation}
where
\begin{align}
a &= \frac{u}{vw} \,, \qquad  b = \frac{v}{wu} \,,
\qquad  c = \frac{w}{uv} \,, \nonumber \\
d &= \frac{1-u}{u} \,, \qquad e = \frac{1-v}{v} \,,
\qquad f = \frac{1-w}{w} \,. \label{abcuvw}
\end{align}
The alphabets $\cL_{u}$ and $\cL_{a}$ are equivalent because they span the same multiplicative space, and the letters in $\cL_{a}$ transform as
\begin{align}
\text{cycle: } &\{a, b, c, d, e, f\} \to \{b, c, a, e, f, d\} \, ,\\
\text{flip: } &\{a, b, c, d, e, f\} \to \{b, a, c, e, d, f\} \, 
\end{align} 
under the generators of the dihedral group.

In refs.~\cite{Dixon:2020bbt,Chicherin:2020umh} it was observed that the set of integrals contributing to form factor at low loop order satisfied a set of adjacency restrictions, which in the alphabet $\cL_{a}$ correspond to the statement that these integrals never involve certain pairs of letters in adjacent slots:
\begin{align}
&\cancel{\dots d \otimes e \dots}\, , \label{eq:adjacency_restrictions_1}
\end{align}
plus the dihedral images of this pair. When $\mathcal{E}^{(L)}$ is expressed in terms of the alphabet $\cL_{a}$, it further becomes apparent from studying the five- and lower-loop results~\cite{Dixon:2020bbt} that its symbol never involves an additional set of pairs:
\begin{align} 
&\cancel{\dots a \otimes d \dots}\, , \qquad \qquad \cancel{\dots d \otimes a \dots}\, , \label{eq:adjacency_restrictions_2}
\end{align}
again plus dihedral images. As these constraints resemble the extended Steinmann relations described in ref.~\cite{Caron-Huot:2019bsq}, we refer to them as ``extended-Steinmann-like'' (ES-like) relations. However, it is worth noting that they do not follow from the standard Steinmann relations~\cite{Steinmann,Steinmann2,Cahill:1973qp} in the form factor space, even in the first two entries of the symbol (which encode the first two discontinuities of $\mathcal{E}^{(L)}$). Indeed, the three-point form factor kinematical variables are all two-particle Mandelstam variables (except for $s_{123}$), and it appears impossible to obtain any Steinmann constraints from discontinuities in such variables for all massless scattering amplitudes. On the other hand, as we will discuss in section~\ref{sec:duality}, these restrictions do in fact follow from the extended Steinmann relations obeyed by the six-point amplitude in planar $\mathcal{N}=4$ sYM theory via the antipodal duality~\cite{Dixon:2021tdw}. Non-trivial Steinmann relations exist for the six-point amplitude because it depends on three independent three-particle invariants.

In fact, we observe additional restrictions on the triple sequences of letters that appear in $\mathcal{E}^{(L)}$, beyond those that are implied by eqs.~\eqref{eq:adjacency_restrictions_1},~\eqref{eq:adjacency_restrictions_2}, and integrability. Namely, the triple
\begin{align} 
&\cancel{\dots a \otimes a b c  \otimes b \dots}\,   \label{eq:triple_adjacency_restriction}
\end{align}
never appears, nor does any of its six dihedral images. (Only four of the six are independent of the pair restrictions, promoted to triple relations.)  These restrictions can be thought of as restrictions on the triple discontinuities of the functions in $\mathcal{C}$. Interestingly, while we do not know how to derive these triple restrictions from form factor physics, they can also be understood from properties of six-point amplitudes using the antipodal duality~\cite{Dixon:2021tdw} (as we discuss further in section~\ref{sec:duality}). The quadruple sequences of letters that appear in the form factor space do not satisfy any restrictions beyond those implied by eqs.~\eqref{eq:adjacency_restrictions_1}, \eqref{eq:adjacency_restrictions_2}, and \eqref{eq:triple_adjacency_restriction}.

Another reason to prefer the functions $\mathcal{E}^{(L)}$ to the remainder function is that they are observed to satisfy powerful multiple-final-entry conditions. In fact, unlike what has been observed in studies of the six- and seven-particle amplitudes in planar $\mathcal{N}=4$ sYM theory, we observe that the function $\mathcal{E}^{(L)}$ satisfies novel multiple-final-entry conditions at each weight. We will discuss these restrictions in detail in section~\ref{sec:final_entry}, and explain them via the antipodal duality in section~\ref{sec:duality}.

The adjacency and multiple-final-entry conditions provide powerful constraints in the bootstrap approach to determining $\mathcal{E}^{(L)}$ that we will describe in the next section. As such, we will bootstrap the function $\mathcal{E}^{(L)}$ directly, rather than the remainder function. Once this function has been determined, the remainder function $R^{(L)}$ can be constructed easily using the relation~\eqref{eq:calE_def}.


\section{The Bootstrap Procedure}
\label{sec: bootstrap procedure}

To compute the form factor $\mathcal{E}$ at higher loops, we follow an improved version of the bootstrap approach used in ref.~\cite{Dixon:2020bbt}, where the form factor was computed at three, four, and five loops.
The $L$-loop function $\mathcal{E}^{(L)}$ is expected to have uniform transcendental weight $2L$. As such, we adopt as our starting ansatz for the function $\mathcal{E}^{(L)}$ a general linear combination of the space of functions whose construction we described below. Namely, we consider the object
\begin{equation} \label{eq:initial_ansatz}
\mathcal{E}^{(L)}=\sum_j c_j F^{(2L)}_j \,,
\end{equation} 
where the sum is over all appropriate functions of weight $2L$, and the coefficients $c_j$ are undetermined rational constants. 
The enlarged set of ES-like adjacency relations~\eqref{eq:adjacency_restrictions_1}, \eqref{eq:adjacency_restrictions_2} and~\eqref{eq:triple_adjacency_restriction} will prevent the dimension of the space from growing too quickly with the weight.  These relations, together with multiple-final-entry conditions and finite field arithmetic to solve the large linear systems of constraints on $c_j$, will enable us to compute the form factor at six, seven, and eight loops.

\subsection{Polylogarithmic Ansatz}
\label{sec:initial_ansatz}

The first step in the bootstrap procedure involves identifying a minimal space of functions within which we expect the form factor to exist at a given loop order. We do this by adopting the conjecture (slightly refined compared to ref.~\cite{Dixon:2020bbt}) that a sufficiently large space of polylogarithms is delineated by the following five properties:  
\begin{enumerate} \itemsep .1cm
\item[(i)] their symbol letters are drawn from the alphabet $\cL_a$, 
\item[(ii)] they are functions, i.e.~they obey integrability relations that follow from the equality of mixed partial derivatives
(see eqs.~\eqref{extrapair1}--\eqref{extrapair3}),
\item[(iii)] they obey the ES-like adjacency relations~\eqref{eq:adjacency_restrictions_1} and~\eqref{eq:adjacency_restrictions_2}, and the triple relations~\eqref{eq:triple_adjacency_restriction},
\item[(iv)] they only develop branch cuts at physical thresholds (at the vanishing loci of $\{a,b,c\}$), 
\item[(v)] the first entry of their coaction only involves polylogarithms that appears in the coaction of the form factor at lower loops.
\end{enumerate} 
The last condition proves extremely useful for getting all the way to eight loops, but requires empirically analyzing the spaces of polylogarithms that appear in the form factor at lower loop orders. We defer its explanation to section~\ref{sec:coaction_principle}, and first proceed by describing the construction of the space defined by properties (i)--(iv). 

The full space of polylogarithms satisfying properties (i)--(iv) can be constructed efficiently by working iteratively in the transcendental weight. Conceptually, this is equivalent to specifying the total derivative of each function in terms of the lower-weight polylogarithms that satisfy the same conditions, supplemented by the choice of an integration constant at a standard base point. However, in practice it is more convenient to phrase these derivatives in the coproduct formalism; for a review of the coproduct structure of polylogarithms, see for instance ref.~\cite{Duhr:2014woa}. This involves describing the $\Delta_{w-1,1}$ coproduct component of each weight-$w$ function, which can be put in the form 
\begin{equation} \label{eq:ansatz_double_coprod_A}
\Delta_{w-1,1} F_{i}^{(w)} =
\sum_{j,k} c_{ijk} F_{j}^{(w-1)} \otimes \ln \phi_k \, ,
\end{equation}
where the sum is over all functions $F_{j}^{(w-1)}$ of weight $w-1$ that satisfy conditions (i)--(iv) and over all letters $\phi_k \in \cL_a$. In the first iteration, when $w=1$, we just need to consider the three weight-one polylogarithms $\{\ln a, \ln b, \ln c\}$ that are allowed by condition (iv).
 
To ensure that the new function $F_{i}^{(w)}$ satisfies property (iii), we replace each
function $\smash{F_j^{(w-1)}}$ by its coproduct component
$\smash{\Delta_{w-2,1} F_j^{(w-1)}}$. This gives us an object of the form
\be
\sum_{j,k} c_{ijk}\big( \Delta_{w-2,1}F_{j}^{(w-1)}  \big) \otimes \ln \phi_k
=  \sum_{j,k,l} c_{ijk} \big(F_{j}^{(w-1)}\big)^{\phi_l} \otimes
\ln \phi_l \otimes \ln \phi_k  \,,
\label{eq:ansatz_double_coprod_B}
\ee 
where we have introduced the notation $F^{\phi_l}$ for the 
`$\phi_l$ coproduct entry' of a function $F$. The ES-like conditions from eqs.~\eqref{eq:adjacency_restrictions_1} and~\eqref{eq:adjacency_restrictions_2} can then be imposed on
the double-coproduct ansatz~\eqref{eq:ansatz_double_coprod_B} by requiring
that the restricted pairs of letters never appear next to each other in the sum over $k$ and $l$:
\begin{align} 
&F^{a, d} = F^{d,a}  = 0 \, , \label{eq:adjacency_constraints_1}   \\ 
&F^{d, e} = 0 \, , \label{eq:adjacency_constraints_2}  
\end{align}
where $F^{\phi_l,\phi_k} \equiv (F^{\phi_k})^{\phi_l}$ is the coproduct entry one obtains by first stripping off the letter $\phi_k$
and then $\phi_l$. When the dihedral images of the restrictions~\eqref{eq:adjacency_constraints_1} and~\eqref{eq:adjacency_constraints_2} are included, one finds 12 independent constraints. 
Since we have started from the artificial object \eqref{eq:ansatz_double_coprod_A}, we must also require (ii) that $F$ is a function, i.e.~that partial derivatives commute, $\partial^2F/(\partial u \partial v) = \partial^2F/(\partial v \partial u)$.  This requirement is known as integrability.  When applied using the first derivative implicit in \eqn{eq:ansatz_double_coprod_A}, the commutator of mixed partial derivatives leads to nine constraints on the antisymmetry combinations $F^{x,y}-F^{y,x}$, where $x,y\in\{a,b,c,d,e,f\}$.  Six of the nine constraints are antisymmetric combinations of \eqns{eq:adjacency_constraints_1}{eq:adjacency_constraints_2}, but there is a further set of three linear conditions,
\bea
&&F^{a,b}+F^{a,c}-F^{b,a}-F^{c,a} = 0, \label{extrapair1}\\
&&F^{c,a}+F^{c,b}-F^{a,c}-F^{b,c} = 0, \label{extrapair2}\\
&&F^{d,b}-F^{d,c}-F^{b,d}+F^{c,d}+F^{e,c}-F^{e,a}-F^{c,e}+F^{a,e}
+F^{f,a}-F^{f,b}-F^{a,f}+F^{b,f} \nonumber\\
&&\hskip0.5cm\null + 4 ( F^{c,b}-F^{b,c} ) = 0, \label{extrapair3} 
\eea
on the coefficients $c_{ijk}$. We also impose the condition~\eqref{eq:triple_adjacency_restriction} on the sequences of triples that appear in the last three entries by imposing all six dihedral images of the constraint
\begin{align} 
&F^{a, a, b} + F^{a, b, b}  + F^{a, c, b} = 0 \, , \label{eq:triple_adjacency_constraints} 
\end{align} 
where $F^{\phi_m,\phi_l,\phi_k} = ((F^{\phi_k})^{\phi_l})^{\phi_m}$. Only four of these constraints are independent after the constraints on pairs of adjacent letters are imposed.

Finally, we must ensure that the branch-cut condition (iv) is obeyed by the function $F_{i}^{(w)}$. This requirement is easiest to state in the alphabet $\cL_u$, and it can be enforced using the conditions
\be
F^{1-u}(1,v,w)\Big|_{v,w\to0} = 0,
\label{branchcutcondition}
\ee
plus the two cyclically related conditions
$F^{1-v}(u,1,w)|_{w,u\to0} = F^{1-w}(u,v,1)|_{u,v\to0} = 0$. 
This condition prevents branch cuts from appearing when one of the letters $u$, $v$, or $w$ approaches $1$. If the left side of eq.~\eqref{branchcutcondition} limits to a transcendental constant, we shift this coproduct entry by the appropriate constant so that this condition is satisfied. However, more generally, the functions we study can become polynomials in $\ln v$ and $\ln w$ 
(with zeta-valued coefficients) in this limit; in these cases, we must remove the offending functions from our basis.

Once we have deduced the full space of coproducts that take the form~\eqref{eq:ansatz_double_coprod_A} and that satisfy conditions (i)--(iv), we set the integration constant by specifying the value of each function at the same kinematic point, $F(1,v\to 0,w \to 0)$. (Concretely, we set as many constants to zero as possible while still preserving dihedral symmetry; see also the discussion in section \ref{subsec:lines}.) This fully specifies the space of functions we are interested in at weight $w$. For a more thorough description of the above-described method, see for instance ref.~\cite{Caron-Huot:2020bkp} (or~ref.~\cite{Dixon:2020bbt} for another discussion specific to $\mathcal{F}_3$).

In practice, this space of functions can also be built -- at least at symbol level -- by imposing conditions iteratively starting from the back of the symbol, rather than the front. Using this method, one begins with an ansatz involving only the symbol letters $d$, $e$, and $f$ in the last entry, due to the last-entry condition that applies in the BDS-like normalization. One then iteratively adds letters to the front of the symbol, using the $\Delta_{1,w-1}$ coproduct, which is analogous to \eqn{eq:ansatz_double_coprod_A} but with $\ln \phi_k$ now appearing in front of the weight $w-1$ symbols. The constraints~\eqref{eq:adjacency_constraints_1} through~\eqref{eq:triple_adjacency_constraints} are applied iteratively in the same way as if one were working from the front. Working from the back, one can make use of the large number of multiple-final-entry constraints that the form factor obeys, which we describe in section~\ref{sec:final_entry}. Conversely, the branch-cut conditions encoded in eq.~\eqref{branchcutcondition} need to be imposed separately at each weight. Even with this drawback, we find that the construction from the back of the symbol proves to be more efficient at high weight than the standard construction that starts with the first entry, due to the large number of multiple-final-entry constraints.

In fact, the most efficient way we have found to proceed at symbol level has been to build the space partially from the back {\it and} from the front. Suppose we find a dimension $f_w$ at weight $w$ for the front space, and a dimension $b_{2L-w}$ at weight $2L-w$ for the back space.  Then we can construct an $L$-loop ansatz with $f_w \times b_{2L-w}$ unknowns, corresponding to the coproduct $\Delta_{w,2L-w}$, where we still need to impose the constraints~\eqref{eq:adjacency_constraints_1} through~\eqref{eq:triple_adjacency_constraints} for pairs and triples of symbol slots that straddle both the front and back spaces.  We call this step {\it sewing across the seam}, and it is the most computationally demanding part of the construction.  After the sewing is completed, the branch-cut condition (iv) still has to be imposed to the right of the seam, followed by conditions discussed further below: dihedral symmetry, the $L^{\rm th}$ discontinuity condition, the leading-power collinear limit, and the near-collinear matching to the FFOPE.

Each of these constraints on the symbol ansatz can be represented as a sparse system of linear equations. Such systems can be solved efficiently over finite fields using the {\sc SpaSM} software library~\cite{spasm}.

After the symbol is determined, terms that are {\it beyond the symbol}, involving zeta values, have to be fixed as well.  However, there are far fewer unknown constants to determine at this stage, so the computational costs associated with the linear algebra are milder.  In this case we work from a complete basis for the function space through weight eight, adding to that the multiple-final-entry conditions described in section~\ref{sec:final_entry}.

\subsection{Summary of Constraints}
\label{sec:summary-constraints}
In order to find the values of the coefficients $c_j$ that correspond to the function $\mathcal{E}^{(L)}$ in eq.~\eqref{eq:initial_ansatz}, we impose various constraints on this ansatz. Namely, we utilize knowledge of various symmetries respected by $\mathcal{E}^{(L)}$, and knowledge of the form it takes in various kinematic limits such as the near-collinear limit. Some of these constraints are easiest to phrase in terms of the remainder function, but imply conditions on $\mathcal{E}$ via the relation~\eqref{eq:calE_def}. They can be summarized as follows:
\begin{enumerate}
\item {\bf Function in $\mathcal{C}$}: We start from the assumption that $\mathcal{E}$ is a function drawn from the form factor space $\mathcal{C}$, which is defined by the conditions (i)-(v) described in the last subsection.
\item {\bf Dihedral symmetry}:  $\mathcal{E}(u,v,w)$ is invariant under any
permutation of $u,v,w$.
\item {\bf Final entry}: The remainder function obeys $\mathcal{E}^{a} = \mathcal{E}^{b} = \mathcal{E}^{c}= 0$,
  resulting in only three linearly independent final entries.
\item {\bf Multiple-final-entries}:  New multiple-final-entry restrictions appear at each weight, which we will describe in section~\ref{sec:final_entry}. These conditions are (partially) built in if we construct from the back.
\item {\bf Collinear limit}: The leading-power collinear limit of $R$
should vanish.
\item {\bf Discontinuity}: The $L^{\rm th}$ discontinuity in $w$ of $R^{(L)}$
should vanish.
\item {\bf Near-collinear limit}:  The near collinear limit of $R$
  should agree with the predictions of the
  FFOPE~\cite{Sever:2020jjx,Sever:2021nsq,Sever:2021xga}.  
\end{enumerate}
As we will show in section~\ref{sec:results}, these requirements are sufficiently constraining to identify a unique function through eight loops, with additional constraints left over as cross-checks. However, before describing the details of the computation at six, seven, and eight loops, we provide additional details on the near-collinear limit, which constitutes an essential source of boundary data.

\subsection{Near-Collinear Limit}
\label{sec:near_coll}

The most important source of boundary data for our bootstrap comes from the near-collinear limit of $R^{(L)}$, which can be computed independently using the FFOPE~\cite{Sever:2020jjx,Sever:2021nsq,Sever:2021xga}. We briefly review this construction here, but refer to~refs.~\cite{Sever:2020jjx,Sever:2021nsq,Sever:2021xga,Dixon:2020bbt} for further details.

The FFOPE directly determines the near-collinear limit of a certain finite ratio of (periodic) Wilson lines $\mathcal{W}_3$. This ratio is related to the remainder function $R$ by 
\begin{equation}
 \label{eq:relation-W-R}
   {\cal W}_3
   =\exp\left[\frac{1}{4}\Gcusp\mathcal{W}^{(1)}_{3}+R\right],
 \end{equation}
 where $\Gcusp$ is the cusp anomalous dimension and
 \begin{equation}
   \mathcal{W}_{3}^{(1)}
   = 4\sigma^2 - 2 \Li_2(- e^{-2\tau}) + 2\Li_2(-e^{-2\tau} - e^{2\sigma})
   + 2 \Li_2(- e^{-2\tau} - e^{-2\sigma}(1 + e^{-2\tau})^2)+\frac{\pi^2}{3}\,.
\label{eq:W31loop}
\end{equation}
 Note that the FFOPE is naturally written in terms of the coordinates $T = e^{-\tau}$ and $S = e^\sigma$, which are related to the ratios of Mandelstam variables in eq.~\eqref{eq:uvwdef} by
 \begin{equation}
 u = \frac{1}{1 + S^2 + T^2} \,, \quad
 v = \frac{T^2}{1 + T^2} \,, \quad 
 w = \frac{1}{(1 + T^2)(1 + S^{-2} (1 + T^2))} \,.
 \label{uvwtoST}
 \end{equation}
The near-collinear expansion corresponds to expanding around small $T$, or equivalently large $\tau$.  We can also write the relation between $\mathcal{W}_{3}$
and $\EE$ directly, using \eqns{eq:E_1loop}{eq:W31loop},
and the kinematical relations~\eqref{uvwtoST}, as
\begin{equation}
 \label{eq:relation-W-E}
   {\cal W}_3
   = \EE \,\exp\left[\frac{1}{4}\Gcusp\mathcal{D}^{(1)}_{3}\right] ,
 \end{equation}
where
\begin{equation}
 \label{eq:Dans}
 \mathcal{D}^{(1)}_{3} \equiv \mathcal{W}^{(1)}_{3} - \EE^{(1)}
  = \ln^2\left(\frac{u}{w}\right) + \ln^2 v
   + 2 \ln(1-v) \ln\left(\frac{u (1-v)}{vw}\right) + 4 \zeta_2 \,.
\end{equation}
Note that the dilogarithms drop out of this relation.

For large $\tau$, the ratio  $\mathcal{W}_{3}$ can be expressed in terms of a sum over eigenstates $\psi$ of the Gubser-Klebanov-Polyakov (GKP) flux tube \cite{Sever:2020jjx}:
 \begin{equation}
   \mathcal{W}_{3}=\int\!\!\!\!\!\!\!\!\!\sum_{\psi}e^{-E_\psi \tau+ip_\psi\sigma}\mathcal{P}(0|\psi)\mathcal{F}(\psi)\,.
 \end{equation}
Exploiting the integrability of the GKP flux tube, it has been possible to determine the energies $E_\psi$ and momenta $p_\psi$ of the states \cite{Basso:2010in}, their pentagon transitions $\mathcal{P}$ \cite{%
   Basso:2013vsa, Basso:2013aha,Basso:2014koa,Basso:2014nra,Basso:2014hfa,%
   Basso:2015rta,Basso:2015uxa,Belitsky:2014sla,Belitsky:2014lta,%
   Belitsky:2016vyq}
 as well as their form factor transitions $\mathcal{F}$ \cite{Sever:2020jjx,Sever:2021nsq,Sever:2021xga} for any value of the planar coupling $g^2$.  Interestingly, the form factor transitions only absorb {\it pairs} of excitations.
 
At any loop order at weak coupling, the terms of order $T^{2n}$ result from states with $2m$ excitations, with $m\leq n$. The momenta of these excitations are integrated over, which can be done via residues to obtain a series expansion in $S^2$.  With some knowledge of the form of the resummed function, and enough terms in the expansion, the full function of $S^2$ can be determined.  The general form of the near-collinear behavior of the form factor remainder function $R^{(L)}$ at $L$ loops is,
\be
R^{(L)}(T,S) = \sum_{j=1}^\infty \sum_{k=0}^{L-1}
  T^{2j} \, \ln^k T \, R_{j,k}^{(L)}(S)  \,.
\label{gennearcol}
\ee
The coefficients $R_{j,k}^{(L)}(S)$ are harmonic polylogarithms
(HPLs)~\cite{Remiddi:1999ew} with indices belonging to $\{0,1\}$ and maximum
weight $2L-k-1$, multiplied by rational prefactors that are universal, i.e.~independent of $L$ and $k$.  The universal prefactors were determined for the $T^2$ and $T^4$ terms in ref.~\cite{Dixon:2020bbt}, and the resummed expressions through $L=5$ loops were given there.  On the other hand, it is not necessary to have the resummed forms to match the boundary data; one can always compare order-by-order in the $S^2$ series.  We followed this approach at higher loop order.  As we will see in section~\ref{sec:imposing_constraints}, the form factor was
determined entirely by the $T^2 \ln^{k}T$ coefficients with $k \geq L-3$,
or next-to-next-to-leading logarithms, and $k \geq L-2$ was sufficient
except at seven loops.


\section{\texorpdfstring{$\mathcal{F}_3$}{F3} at Six, Seven, and Eight Loops}
\label{sec:results}

In order to compute $\mathcal{F}_3$ beyond five loops, we start from the ansatz described in section~\ref{sec:initial_ansatz}, where we now incorporate property (v) into its construction.  Property (v) corresponds to including information about the minimal space of polylogarithms required to represent the form factor at lower loops, which can be formalized as a Galois coaction principle, similar to what has been found in other contexts~\cite{Schlotterer:2012ny,Schnetz:2013hqa,Brown:2015fyf,Panzer:2016snt,Schnetz:2017bko,Caron-Huot:2019bsq,Gurdogan:2020ppd}. Adopting the notation of~ref.~\cite{Dixon:2020bbt}, we call the minimal space of functions that contributes to $\mathcal{F}_3^{(L)}$ the $\mathcal{C}$ space. However, note that we have refined the $\mathcal{C}$ space using the additional data bootstrapped in this paper, and so its dimension has decreased relative to the space of the same name defined in ref.~\cite{Dixon:2020bbt}.

In the process of bootstrapping $\mathcal{F}_3$ to higher loops, we have also observed new multiple-final-entry conditions at each weight, counting from the back of the symbol. In fact, we find that the form factor saturates the space allowed by these multiple-final-entry conditions at one higher weight at each loop order. This has allowed us to identify (at seven loops, and verify at eight loops) a minimal space of 155 heptuple final entries that are consistent with the quadruple-final-entry space identified in ref.~\cite{Dixon:2020bbt}.

While both the Galois coaction principle and the multiple-final-entry conditions we observe are empirical, they can be checked iteratively in the loop order and provide important constraints on our ansatz at higher loops. We thus describe how we observe both of these sets of constraints before presenting the details of the six-, seven-, and eight-loop bootstrap computation.

\subsection{Galois Coaction Principle}
\label{sec:coaction_principle}

\renewcommand{\arraystretch}{1.25}
\begin{table}[!t]
\centering
\begin{tabular}[t]{l c c c c c c c c}
\hline\hline
$w$                       &  1  &  2 &  3  &  4  &  5  &  6  &  7  &  8
\\\hline\hline
${\cal M}$ (\eqref{eq:adjacency_restrictions_1} only)
                          &  3  &  9 &  27 & 81 & 243 & 729 & 2187 & 6561
\\\hline
add \eqref{eq:adjacency_restrictions_2}
and \eqref{eq:triple_adjacency_restriction}
                          &  3  &  9 &  21 & 51 & 120 & 279 & 642 & 1470
\\\hline
${\cal C}$ symbols        & 3   &  9 &  21 & 48 & 108 & 249 & 567 & 1290
\\\hline
${\cal C}$ functions      & 3   &  9 &  22 & 52 & 122 & 284 & 654 & 1495
\\\hline
\end{tabular}
\caption{Dimension of various possible polylogarithmic spaces, as a function
of the weight $w$.  Except for the last row, the numbers are at symbol level.
We remark that 
when lifting ${\cal C}$ from symbol to function level, a few symbols have
to be removed, starting at weight 6; if one were to compute the symbol of the $\mathcal{C}$ function space, one would find only 
246, 555, 1251 independent symbols  at weights 6, 7, 8.}
\label{tab:variousspaces}
\end{table}


As was already observed in ref.~\cite{Dixon:2020bbt}, the perturbative expansion of $\mathcal{E}$ does not make use of the full space of polylogarithms defined by properties (i)--(iv) in section~\ref{sec:initial_ansatz}.  First of all, if we had not used the pair and triple conditions~\eqref{eq:adjacency_restrictions_2} and \eqref{eq:triple_adjacency_restriction}, we would have arrived at the space ${\cal M}$ defined in ref.~\cite{Dixon:2020bbt}, which at symbol level has dimension $3^w$ at weight $w$; see Table~\ref{tab:variousspaces}.  Imposing the pair and triple conditions~\eqref{eq:adjacency_restrictions_2} and \eqref{eq:triple_adjacency_restriction} leads to the much smaller dimensions in the second row.

\renewcommand{\arraystretch}{1.25}
\begin{table}[!t]
\centering
\begin{tabular}[t]{l c c c c c c c c c c c c c c c c c}
\hline\hline
weight $n$
& 0 & 1 & 2 & 3 & 4 &  5 &  6 &  7 &  8 &  9 & 10 & 11 & 12 & 13 & 14 & 15 & 16
\\\hline\hline
$L=1$
& \green{1} & \blue{3} & \blue{1} &  &  &  &  &  &  &  & & & & & & &
\\\hline
$L=2$
& \green{1} & \green{3} & \blue{6} & \blue{3} & \blue{1}
& & & & & & & & & & & &
\\\hline
$L=3$
& \green{1} & \green{3} & \green{9} & \blue{12} & \blue{6} & \blue{3}
& \blue{1} &  &  &  & & & & & & &
\\\hline
$L=4$
& \green{1} & \green{3} & \green{9} & \green{21} & \blue{24} & \blue{12}
& \blue{6} & \blue{3} & \blue{1} &  & & & & & & &
\\\hline
$L=5$
& \green{1} & \green{3} & \green{9} & \green{21} &  46 & \blue{45}
& \blue{24} & \blue{12} & \blue{6} & \blue{3} & \blue{1} & & & & & &
\\\hline
$L=6$
& \green{1} & \green{3} & \green{9} & \green{21} & \green{48} & 99
& \blue{85} & \blue{45} & \blue{24} & \blue{12} & \blue{6} & \blue{3} & \blue{1}
& & & &
\\\hline
$L=7$
& \green{1} & \green{3} & \green{9} & \green{21} & \green{48} & \green{108}
& 236 & \blue{155}
& \blue{85} & \blue{45} & \blue{24} & \blue{12} & \blue{6} & \blue{3} & \blue{1}
& &
\\\hline
$L=8$
& \green{1} & \green{3} & \green{9} & \green{21} & \green{48} & \green{108}
& 242 & 466 & \red{279} & \blue{155}
& \blue{85} & \blue{45} & \blue{24} & \blue{12} & \blue{6} & \blue{3} & \blue{1}
\\\hline\hline
\end{tabular}
\caption{The number of independent $\{n,1,1,\ldots,1\}$ coproducts
  of the form factor $\EE^{(L)}$ through $L=8$ loops, \emph{at symbol level}.
  A green number denotes saturation of the space $\cC$ constructed from
  the bottom up. A blue number denotes saturation of the space of coproducts,
  as indicated by the $(L,n)$ number being the same as the $(L+1,n+2)$ number.
  A red number means it is almost certainly saturated, but we do not have
  the evidence of two successive loop orders to ``prove'' it yet.}
\label{tab:EEcopdimsymb}
\end{table}

The dimensions in  Table~\ref{tab:variousspaces} can be compared to the entries of Table~\ref{tab:EEcopdimsymb}, where we present the number of independent $\{n,1,1,\ldots,1\}$ coproduct entries that we will find at symbol level in our eventual answer for $\mathcal{E}^{(L)}$. For instance, the dimension of the space of weight-four symbols that appear in the first entry of $\mathcal{E}^{(L)}$ through eight loops is just 48. This is much smaller than the 81 symbols in ${\cal M}$, but it is also three smaller than the 51 symbols obtained by building the full space of symbols consistent with properties (i)--(iv).  We identify and remove the extra three symbols in defining ${\cal C}$ at symbol level.  We then find 108 symbols at weight 5, which matches the number observed in Table~\ref{tab:EEcopdimsymb}.

\renewcommand{\arraystretch}{1.25}
\begin{table}[!t]
\centering
\begin{tabular}[t]{l c c c c c c c c c c c c c c c c c}
\hline\hline
weight $n$
& 0 & 1 & 2 & 3 & 4 &  5 &  6 &  7 &  8 &  9 & 10 & 11 & 12 & 13 & 14 & 15 & 16
\\\hline\hline
$L=1$
& \green{1} & \blue{3} & \blue{1} &  &  &  &  &  &  &  & & & & & & &
\\\hline
$L=2$
& \green{1} & \green{3} & \blue{6} & \blue{3} & \blue{1} &
& & & & & & & & & & &
\\\hline
$L=3$
& \green{1} & \green{3} & \green{9} & \blue{12} & \blue{6} & \blue{3}
& \blue{1} &  &  &  & & & & & & &
\\\hline
$L=4$
& \green{1} & \green{3} & \green{9} & 21 & \blue{24} & \blue{12}
& \blue{6} & \blue{3} & \blue{1} &  & & & & & & &
\\\hline
$L=5$
& \green{1} & \green{3} & \green{9} & 21 &  47 & \blue{45}
& \blue{24} & \blue{12} & \blue{6} & \blue{3} & \blue{1} & & & & & &
\\\hline
$L=6$
& \green{1} & \green{3} & \green{9} & \green{22} & 49 & 103
& \blue{85} & \blue{45} & \blue{24} & \blue{12} & \blue{6} & \blue{3} & \blue{1}
& & & &
\\\hline
$L=7$
& \green{1} & \green{3} & \green{9} & \green{22} & \green{51} & \green{115}
& 243 & \blue{155}
& \blue{85} & \blue{45} & \blue{24} & \blue{12} & \blue{6} & \blue{3} & \blue{1}
& &
\\\hline
$L=8$
& \green{1} & \green{3} & \green{9} & \green{22} & \green{51} & \green{115}
& 258 & 466 & \red{279} & \blue{155}
& \blue{85} & \blue{45} & \blue{24} & \blue{12} & \blue{6} & \blue{3} & \blue{1}
\\\hline\hline
\end{tabular}
\caption{The number of independent $\{n,1,1,\ldots,1\}$ coproducts
  of the form factor $\EE^{(L)}$ through $L=8$ loops, \emph{at function level}.
  The colors have the same meaning as in Table~\ref{tab:EEcopdimsymb}.}
\label{tab:EEcopdim}
\end{table}

The same observation can be extended to the level of full functions, by analyzing the first entry of the coaction of $\mathcal{E}^{(L)}$ at all available loop orders. Namely, the span of the functions of fixed weight $w$ that appear in the first entry of the functions $\mathcal{E}^{(L)}$ stabilizes after a certain number of loop orders. After this stabilization has been observed, it restricts the space of weight-$w$ functions we expect to appear in the first entry of the coaction of $\mathcal{E}^{(L)}$ at higher loop orders. Formally, this restriction can be expressed as a coaction principle, namely
\begin{equation}
\Delta \mathcal{C} \subset \mathcal{C} \otimes \mathcal{K} \, , \label{eq:coaction_principle}
\end{equation}
where $\mathcal{K}$ is the space of de Rham functions (or polylogarithms modulo $i \pi$) that are consistent with properties (i)--(iii). In other words, the coaction principle~\eqref{eq:coaction_principle} states that the space of functions $\mathcal{C}$, which by definition contains $\mathcal{E}^{(L)}$ to all loop orders, can be constructed at the level of the coaction using knowledge of the span of $\mathcal{C}$ at lower weights. Similar coaction principles have been observed in other perturbative contexts~\cite{Brown:2015fyf,Panzer:2016snt,Schnetz:2017bko,Caron-Huot:2019bsq}.

While we trust the restrictions we get from these observations, we do not yet have a precise understanding of when we expect $\mathcal{C}$ to stabilize for a given weight. In practice, we observe that these low-weight spaces continue to grow at symbol level at every loop order until saturation; thus, once the space of symbols appearing in the first entry of $\mathcal{E}^{(L)}$ is observed to have the same dimension at two successive loop orders, we adopt this as the definition of $\mathcal{C}$ at that weight. As can be seen at weight 3 in Table~\ref{tab:EEcopdim}, at function level we must be more cautious about when we take $\mathcal{C}$ to be saturated, since this space was observed to have dimension 21 twice, before saturating at 22. Above the weights where we believe saturation has been achieved, we continue to work with the full space of functions that is consistent with properties (i)--(iv), but whose first coproduct entry can be expressed in terms of this smaller (saturated) space of functions. 

By analyzing the space of functions that appear in $\mathcal{E}$ through eight loops, we find that there exists a closely-related normalization that involves a reduced space of transcendental constants at certain kinematic points. This normalization is defined by
\begin{equation}
\mathcal{E}_c =  \frac{ {\cal E} }{\rho_3} = F_3 \, ,
\label{eq:EEcrelation}
\end{equation}
where $F_3$ is the notation used in ref.~\cite{Dixon:2021tdw},
\begin{equation}
\rho_3 = \det(1 + \mathbb{K}) e^{-\zeta_2 \Gcusp} \, 
\end{equation}
and $\mathbb{K}$ is the BES kernel~\cite{Beisert:2006ez},
as defined e.g.~in ref.~\cite{Basso:2020xts}.
Through six loops, $\rho_3$ is given by
\begin{align}
\rho_3 &= 1 - 2 \zeta_2 g^2
+ 14 \zeta_4 g^4
- \left(\frac{293}{3} \zeta_6 - 8 \zeta_3^2 \right) g^6 + \left(\frac{7148}{9}\zeta_8-160 \zeta_3 \zeta_5 \right) g^8 \nonumber \\
&\qquad 
- \left(160 \zeta_4 \zeta_3^2 + \frac{21371}{3} \zeta_{10} - 912 \zeta_5^2  - 1680 \zeta_3 \zeta_7 \right) g^{10}  + \bigg(\frac{6080}{3} \zeta_6 \zeta_3^2 \\
&\qquad
-192 \zeta_2 \zeta_5^2
  +3008 \zeta_4 \zeta_3 \zeta_5-20832 \zeta_5 \zeta_7-18816 \zeta_3\zeta_9
  +\frac{419107544}{6219}\zeta_{12}\bigg)  g^{12} \, . \nonumber
\end{align}
Its value through eight loops is given in the ancillary file {\tt AntipodePointsSummary.txt} for ref.~\cite{Dixon:2021tdw}.
 
It is important to note that in contrast to $\mathcal{E}$, the function $\mathcal{E}_c$ is not in the space $\mathcal{C}$, because the $\zeta_2$ shift at weight two cannot be accommodated. This is due to the fact that $\zeta_2$ is not an independent element of $\mathcal{C}$; it is always linked to the nine weight-two functions indicated in Table \ref{tab:EEcopdim}.  In order to work with the smallest function space possible, we bootstrap the function $\mathcal{E} \in \mathcal{C}$, and compute $\mathcal{E}_c$ from \eqn{eq:EEcrelation}, only after computing $\mathcal{E}$.

To describe the special behavior exhibited by the function $\mathcal{E}_c$, we consider its value at the point where $u$ and $v$ are both sent to positive infinity. At this point, both the functions $\mathcal{E}$ and $\mathcal{E}_c$ are finite and can be expressed in terms of multiple zeta values (MZVs). However, in contrast to $\mathcal{E}^{(L)}(\infty,\infty,-\infty)$, the number $\mathcal{E}_c^{(L)}(\infty,\infty,-\infty)$ has the additional property that its coproduct never involves $\zeta_3$ in the right entry. In the language of `$f$-alphabets'~\cite{Brown:2011ik} this corresponds to the statement that the derivation $\partial_3$, which picks out the component of the coproduct whose right entry is $\zeta_3$, annihilates the form factor:
\begin{equation}
\partial_3 \mathcal{E}_c^{(L)}(\infty,\infty,-\infty) = 0 \, . \label{eq:partial_3_property}
\end{equation}
This property is easy to see in the analytic values we present for $\mathcal{E}_c^{(L)}(\infty, \infty, - \infty)$ in section~\ref{sec:special_points}; in our conventions, it corresponds to the fact that none of the $f$-alphabet words that appear start with an index of $3$. It would be interesting to understand the physical interpretation of the constraint~\eqref{eq:partial_3_property}.

Importantly, it is the function $\mathcal{E}_c$ that enters the antipodal duality between the form factor and the six-point MHV amplitude. In particular, while $\mathcal{E}$ and $\mathcal{E}_c$ are identical at symbol level, only the latter function correctly recaptures the transcendental constants (other than $i \pi$, whose antipode is not defined) that appear in the six-point amplitude on the other side of the duality. 
As the antipode has the effect of reversing the words in our $f$-alphabet (up to a sign), the constraint~\eqref{eq:partial_3_property} implies that $\zeta_3$ should not be observed in the first entry of the six-point amplitude at the corresponding point $(\hat{u},\hat{v},\hat{w}) = (1,1,1)$ in amplitude kinematics, to any loop order. This is precisely the property that was observed in ref.~\cite{Caron-Huot:2019bsq}.  Furthermore, the $\mathcal{E}_c$ normalization improves the antipodal duality properties in the Euclidean soft and collinear limits, where $\mathcal{E}_c$ matches precisely the dual collinear and soft limits of the cosmically normalized six-point MHV amplitude.



\subsection{Multiple-Final-Entry Conditions}
\label{sec:final_entry}

At the right end of the symbol, we also observe an increasingly powerful set of multiple-final-entry constraints that are obeyed by $\mathcal{E}^{(L)}$. The single-, double-, and triple-final-entry constraints obeyed by these functions were already characterized in~ref.~\cite{Dixon:2020bbt}, but with data through eight loops we can now describe with confidence a set of quadruple-, pentuple-, hextuple-, and heptuple-final-entry constraints. Unlike the first entries of the symbol, this final-entry space seems to saturate at a predictable rate: we saturate the space of final entries at one higher weight with each loop order, starting with the single-final-entry space at one loop.  In other words, once we have the $L$-loop answer, we can predict all relations between the final $L$ entries.

As stated above, the single-final-entry conditions on the BDS-like normalized functions are given by
\begin{align}
\mathcal{E}^a = \mathcal{E}^b = \mathcal{E}^c = 0 \, . \label{eq:final_entry_again}
\end{align} 
(Both here and below, we phrase coproduct-level constraints in terms of the finite-coupling form factor $\mathcal{E}$, although in practice we always impose them at a fixed loop order $\mathcal{E}^{(L)}$.) Once the final-entry conditions~\eqref{eq:final_entry_again} are imposed, we find that additional double-final-entry conditions are satisfied by $\mathcal{E}$. These can be phrased as the constraints 
\begin{align}
\mathcal{E}^{a,d} = \mathcal{E}^{e,d} = 0 \, , \qquad \mathcal{E}^{b,f} = \mathcal{E}^{b,d} \, ,
\end{align} 
as well as their dihedral images. This leaves only six independent double-final coproduct entries. Three entries from the end of the symbol, we yet again observe new conditions that go beyond the double- and single-final-entry conditions. These new conditions take the form
\begin{align}
\mathcal{E}^{c,d,d} = -\mathcal{E}^{c,e,e} \, , \qquad \mathcal{E}^{f,b,d} = \mathcal{E}^{d,b,d} - \mathcal{E}^{b,d,d}  \, , \\
\mathcal{E}^{a,d,d} = \mathcal{E}^{a,b,d} = \mathcal{E}^{a,c,e} = \mathcal{E}^{e,b,d} = \mathcal{E}^{e,d,d} = 0\, ,
\end{align}
plus all dihedral images. In combination with the constraints from a lower number of final entries, these leave twelve independent triple-final-entry coproducts. Four entries from the end of the symbol, we again find new conditions that are not implied by the previous final-entry conditions:
\begin{gather}
\mathcal{E}^{a,d,d,d} = \mathcal{E}^{a,b,b,d} = \mathcal{E}^{a,d,b,d} = \mathcal{E}^{c,b,b,d} = 0 \, , \\
\mathcal{E}^{e,b,b,d} = \mathcal{E}^{e,b,d,d} = \mathcal{E}^{e,d,b,d} = \mathcal{E}^{e,d,d,d} = \mathcal{E}^{f,d,b,d} = 0 \, , \\
\mathcal{E}^{d,d,b,d} = \mathcal{E}^{d,b,d,d} \, , \\
\mathcal{E}^{b,d,d,d} = -\mathcal{E}^{f,a,f,f}+\mathcal{E}^{d,b,d,d}+\mathcal{E}^{e,a,f,f}-\mathcal{E}^{f,b,d,d}+\mathcal{E}^{a,e,e,e} \, , \\
\mathcal{E}^{a,b,d,d} = \frac{1}{4} \left(\mathcal{E}^{c,d,d,d}+ \mathcal{E}^{d,c,e,e}- \mathcal{E}^{a,e,e,e}-\mathcal{E}^{e,a,f,f}+ \mathcal{E}^{f,a,f,f}-\mathcal{E}^{e,c,e,e} \right) \, , \\
\mathcal{E}^{c,b,d,d} = \mathcal{E}^{c,d,b,d} = \frac{1}{4}  \left( \mathcal{E}^{b,f,f,f}  -\mathcal{E}^{d,c,e,e}+ \mathcal{E}^{e,c,e,e}- \mathcal{E}^{c,d,d,d}- \mathcal{E}^{d,b,d,d}+ \mathcal{E}^{f,b,d,d}\right) \, , \\
\mathcal{E}^{f,b,b,d} = \mathcal{E}^{d,b,b,d} -\mathcal{E}^{b,b,d,d} + \frac{1}{4} \left( \mathcal{E}^{f,a,f,f} {-} \mathcal{E}^{d,b,d,d} {+} \mathcal{E}^{f,b,d,d} {-} \mathcal{E}^{e,a,f,f} {-} \mathcal{E}^{a,e,e,e} {+} \mathcal{E}^{b,f,f,f} \right) \, , \label{eq:last_quad_final}
\end{gather}
plus all dihedral images. These equations leave 24 independent quadruple-final-entry coproducts.  The number of relations required through weight 8 is shown in Table~\ref{tab:finalentrycounting}.  We record the higher-weight multiple-final-entry conditions, which are too large and numerous to include here, in an ancillary file, {\tt FFmultifinalentry.txt}. 

\renewcommand{\arraystretch}{1.25}
\begin{table}[!t]
\centering
\begin{tabular}[t]{l c c c c c c c c}
\hline\hline
$w$                                     &  1  &  2 &  3  &  4  &  5  &  6  &  7  &  8
\\\hline\hline
na\"ive number of final $w$ entries
                                 & 6 & 18 & 36 & 72 & 144 & 270 & 510 & 930
\\\hline
number of relations              & 3 & 12 & 24 & 48 & 99 & 185 & 355 & 651
\\\hline
after final $w$ entry conditions & 3 &  6 & 12 & 24 & 45 &  85 & 155 & 279  \\\hline
\end{tabular}
\caption{Before imposing the final $w$ entry relations, the ``na\"ive'' number of final $w$ entries is 6 times the actual number of final entries at weight $w-1$, because one can put any of the 6 symbol letters in front of an independent $w-1$ final entry.  The number of relations required is the difference between this number and the actual number of final entries at weight $w$.}
\label{tab:finalentrycounting}
\end{table}

Interestingly, the dimensions of these final-entry spaces do not seem to follow a simple pattern. At weights one through eight (counting from the back of the symbol), they can be seen from Table~\ref{tab:EEcopdimsymb} to have dimension $\{ 3, 6, 12, 24, 45, 85, 155, 279\}$. The number of new constraints at each higher weight also does not seem to follow a clear pattern. In Table~\ref{tab:finalentryconditions}, we record both the dimension of the weight-$w$ space of final entries that follows from the weight-$(w-1)$ final-entry conditions when combined with the pair and triplet constraints (as given in eqs.~\eqref{eq:adjacency_constraints_1}, \eqref{eq:adjacency_constraints_2}, and \eqref{eq:triple_adjacency_constraints}), and the actual dimension of the weight-$w$ final entry space. (Note that eqs.~\eqref{eq:final_entry_again}--\eqref{eq:last_quad_final} at weight $w-1$ subsume the generic pair and triple constraints at weight $w-1$, but they do not include the generic pair and triple constraints acting at weight $w$.) It would be interesting to be able to predict these constraints at higher weight, in order to be able to use them before first observing them at the previous loop order. As we will discuss in section \ref{sec:duality}, this is indeed possible; by antipodal duality, they are related to the initial entries for the space of hexagon functions (describing six-point scattering amplitudes) on the surface $\Delta=0$, which is better understood.

\renewcommand{\arraystretch}{1.25}
\begin{table}[!t]
\centering
\begin{tabular}[t]{l c c c c c c c c}
\hline\hline
$w$                                     &  1  &  2 &  3  &  4  &  5  &  6  &  7  &  8
\\\hline\hline
before final $w$ entry conditions       &  6  &  8 &  15 &  26 &  49 &  87 & 164 & 289
\\\hline
after final $w$ entry conditions        &  3  &  6 &  12 &  24 &  45 &  85 & 155 & 279
\\\hline
new final $w$ entry conditions      &  3  &  2 &   3  &  2  &  4 &   2 &   9 & 10 \\\hline
\end{tabular}
\caption{Dimension of the space of potential $\{2L-w,1,\dots,1\}$ $w^\text{th}$ coproduct entries of the $L$-loop form factor, before and after imposing the $w$-final-entry conditions. We also give their difference, which is the number of non-trivial, new constraints at weight $w$ (not accounting for dihedral symmetry).}
\label{tab:finalentryconditions}
\end{table}

\subsection{Imposing the Constraints}
\label{sec:imposing_constraints}

Putting all of the above ingredients together, we have bootstrapped $\mathcal{E}^{(L)}$ at six, seven, and eight loops. With the computation of each successive loop order, we have been able to refine our best guess for the minimal space $\mathcal{C}$. As can be seen in Tables~\ref{tab:EEcopdimsymb} and~\ref{tab:EEcopdim}, only the spaces in $\mathcal{C}$ with weight up to five have saturated by eight loops.

\renewcommand{\arraystretch}{1.25}
\begin{table}[!t]
\centering
\begin{tabular}[t]{l c c c c c c c}
\hline\hline
$L$                        &  2 &   3 &   4  &   5  &  6  &  7  &  8
\\\hline\hline
symbols in $\cC$           & 48 & 249 & 1290 & 6654 & 34219 & ???? & ????
\\\hline
dihedral symmetry          & 11 &  51 &  247 & 1219 & ???? & ???? & ????
\\\hline
$(L-1)$ final entries      &  5 &   9 &   20 &  44 &  86 & 191 & 191\\\hline
$L^{\rm th}$ discontinuity   &  2 &   5 &  17 &   38 &  75 & 171 & 164 \\\hline
collinear limit            &  0 &   1 &    2 &   8 &  19 & 70 & 6 \\\hline
OPE $T^2\,\ln^{L-1}T$       &  0 &   0 &    0 &   4 &  12 & 56 & 0 \\\hline
OPE $T^2\,\ln^{L-2}T$       &  0 &   0 &    0 &   0 &   0 & 36 & 0 \\\hline
OPE $T^2\,\ln^{L-3}T$       &  0 &   0 &    0 &   0 &   0 & 0 & 0 \\\hline
OPE $T^2\,\ln^{L-4}T$       &  0 &   0 &    0 &   0 &   0 & 0 & 0 \\\hline
OPE $T^2\,\ln^{L-5}T$       &  0 &   0 &    0 &   0 &   0 & 0 & 0 \\\hline\hline
\end{tabular}
\caption{Number of parameters left when bootstrapping the form factor
  $\EE^{(L)}$ at $L$-loop order in the function space $\cC$
  at symbol level, using all the conditions on the final $(L-1)$ entries,
  which can be deduced at $(L-1)$ loops.}
\label{tab:Cparameters}
\end{table}

In Table~\ref{tab:Cparameters}, we present the number of free coefficients left in our ansatz at each stage in the bootstrap procedure. To make these numbers comparable across loop orders, we have adopted the same guess for the minimal space $\mathcal{C}$ in the initial step, which assumes knowledge of its saturation at weight four.
As mentioned previously, we did not construct the full space $\mathcal{C}$ from the front past weight eight. We found it more efficient to build it from the back as well as from the front, and sew the two spaces together by imposing the remaining conditions across the seam. The $(L-1)$-final-entry conditions are built into the construction of the back-end space. For this reason, we do not know how many independent symbols there are in $\mathcal{C}$ past weight 12, accounting for the question marks in the top row of Table \ref{tab:Cparameters}. Similarly, we do not know how many of the 34219 symbols in $\mathcal{C}$ at weight 12 are dihedrally symmetric, accounting for the question marks in the next row.

We see from Table \ref{tab:Cparameters} that there are only a modest number of parameters left by the time the $(L-1)$ final entry conditions are imposed.  The imposition of the $L^{\rm th}$ discontinuity condition was discussed in ref.~\cite{Dixon:2020bbt}, albeit in a different alphabet and for a different normalization of the form factor. We impose the vanishing of the $L^{\rm th}$ discontinuity in $w$ for the remainder function $R$. We have to clip the symbol
$L$ times from the left with the operator
\be
\disc_w = - \disc_a - \disc_b + \disc_c - \disc_f \,,
\label{eq:Cw}
\ee
which follows from how $w$ appears in the letter definitions~(\ref{abcuvw}).
At symbol level, using \eqn{eq:calE_def}, we impose
\be
0 = (\disc_w)^L R^{(L)} =  (\disc_w)^L \big[ \ln \EE \big]_{g^{2L}} \,.
\label{eq:Ldisc}
\ee
We expand the right-hand side perturbatively and make use of the
Leibniz rule for discontinuities of products, as well as knowledge of the
lower-loop $\EE^{(\ell)}$ symbols.

The form-factor normalization used previously~\cite{Dixon:2020bbt}
had the property that the second discontinuity
of the one-loop amplitude vanished, which in turn implied that only
the term with $L$ powers of the one-loop amplitude contributed
in the relation analogous to \eqn{eq:Ldisc}.  In the present normalization,
at symbol level, we see from \eqn{eq:Cw} that the second discontinuity
is non-vanishing:
\bea
\mathcal{S} \Bigl[ \EE^{(1)} \Bigr]
&=& - a\otimes e - a\otimes f - b\otimes f - b\otimes d
             - c\otimes d - c\otimes e \,,
\label{E1}\\
\mathcal{S} \Bigl[ \disc_w \, \EE^{(1)} \Bigr] &=& 2 \times f \,,
\label{DwE1}\\
\mathcal{S} \Bigl[ (\disc_w)^2 \, \EE^{(1)} \Bigr] &=& - 2 \,.
\eea
The consequence is that imposing condition~\eqref{eq:Ldisc} requires knowing (all discontinuities up to) the $L^{\rm th}$ discontinuities of all lower-loop symbols $\mathcal{E}^{(L-i)}$, as well as shuffling many of these symbols together in order to obtain the relevant contributions.  Imposing the condition on the ansatz at eight loops requires computing $(\disc_w)^8$ on all the weight 8 symbols in $\mathcal{C}$, which one can work up to from lower weights, using the fact that discontinuities and derivatives commute. In the end, a relatively small number of parameters are fixed by condition~(\ref{eq:Ldisc}).

The leading-power (or strict) collinear limit is considerably more powerful
and straightforward to impose. At symbol level, from \eqn{eq:calE_def} and that fact that $R \to 0$ in this limit, we find that
\be
\EE^{(L)}\bigr|_{v\to0} = \frac{1}{L!} \Bigl[ \EE^{(1)}\bigr|_{v\to0} \Bigr]^L
= \frac{1}{L!} ( - \ln^2 v - \ln^2 d )^L \,.
\label{eq:strictcoll}
\ee
The form factor ansatz in the strict collinear limit collapses to
HPLs with indices belonging to $\{0,1\}$, so even the function-level
version of this constraint is easy to impose.  At function level,
one does have to fix the constants of integration at the base point
$(u,v) = (1,0)$, after defining all the basis functions there.

After ensuring that the remainder function $R^{(L)}$ vanishes in the strict
collinear limit, the final step is to match the FFOPE
predictions~\cite{Sever:2020jjx,Sever:2021nsq,Sever:2021xga}
for the near-collinear limit at order $T^2 \approx v^1$.
One can do so by integrating up a differential equation in $v$, which needs
no boundary condition beyond $v^0$ (which corresponds to the strict-collinear limit).
Table \ref{tab:Cparameters} shows that all parameters are fixed
at symbol level by order $T^2 \, \ln^{L-3} T$ at seven loops,
and even earlier at other loop orders.  They are also fixed
fairly quickly at function level.  Further subleading orders in $\ln T$,
beyond the first zero in each column, then serve as unambiguous cross checks
of the result, which are successfully passed. (Concretely, we checked the terms of order $T^2 \ln^k T$ for all $k = 0,1,2,\ldots,L-1$ in a series expansion in small $S$ up to at least order $S^{80}$ at six loops, at least order $S^{60}$ at seven loops and at least order $S^{20}$ at eight loops, checking also much higher orders in $S$ for $k>0$.)
Most strikingly, we see there that more FFOPE data is needed to bootstrap $\mathcal{E}^{(L)}$ at seven loops than at eight loops. We suspect that the reduction in the required OPE data at eight loops is caused by the large number of independent multiple-final-entry conditions which we observe at seven loops and first use at eight loops (cf.~Table \ref{tab:finalentryconditions}).
Since we observe a similarly-large number of independent multiple-final-entry conditions at eight loops,
we expect that the amount of OPE data required at nine loops is equally
limited.

In the ancillary files {\tt T2terms8loops.txt} and {\tt T4terms8loops.txt},
we provide the near-collinear expansion of the remainder function $R^{(L)}$
through $L=8$ at orders $T^2$ and $T^4$, respectively.  We use a different
notation from ref.~\cite{Dixon:2020bbt}, featuring HPLs
with argument $y = S^2 \approx (1-u)/u$, fully linearized.
The $T^4$ terms were also checked against the FFOPE
predictions~\cite{Sever:2020jjx,Sever:2021nsq,Sever:2021xga}
at order $T^4 \ln^k T$, for all non-trivial terms, $k = 0,1,2,\ldots,L-1$,
through five loops, in an expansion for small $S$ through order $S^{10}$.
In this paper, they were similarly checked all the way through eight loops for
$k = L-3,\,L-2,\,L-1$, as well as for $k=2$ at six loops.
We provide the series expansions in $S$ through order $S^{20}$ in
the ancillary files {\tt T2termsyser8loops.txt}
and {\tt T4termsyser8loops.txt}.

In the ancillary file {\tt Esymb.txt}, we provide the symbol of the form factor through seven loops.  The seven-loop symbol would be many gigabytes
if fully expanded out, so we give the results for a set of 93 independent
octuple final entries (there are 279 linearly independent octuple final
entries, which can be arranged into 93 three-orbits under the cyclic
transformation).  At intermediate loop orders, we also give the symbol
in terms of 8 independent three-orbits of quadruple final entries.
Even with this ``octuple compression'', the eight-loop symbol is 154MB,
and it is provided in a separate file, {\tt Esymboct8.txt}.

We also give the full function-level result for $\mathcal{E}^{(L)}$ through eight loops in the file {\tt E\_C.txt}. 
This file makes use of the $\mathcal{C}$ space of functions through weight eight, which is itself given in
the files {\tt Ccop8.txt} and {\tt Cfns.txt}. Above this weight, we must also specify the value of the 
independent coproducts of $\mathcal{E}^{(L)}$ at the point $(1,0,0)$. These are given, alongside the
values of the full function $\mathcal{E}^{(L)}(1,0,0)$, in the file {\tt E100.txt}.


\section{Consequences of Antipodal Duality}
\label{sec:duality}

In a recent letter \cite{Dixon:2021tdw}, the authors reported a
duality in planar $\mathcal{N}=4$ sYM theory between the cosmically normalized MHV three-point form
factor $\mathcal{E}_c$, and the cosmically normalized MHV six-point amplitude $\mathcal{E}_6$.\footnote{In ref.~\cite{Dixon:2021tdw}, $\mathcal{E}_6$ was called $A_6$, and $\mathcal{E}_c$ was called $F_3$.}
According to this duality, the form factor is related by the antipode map (which forms part of the Hopf algebra defined on multiple polylogarithms~\cite{Gonch3,Gonch2,Goncharov:2010jf,Brown:2011ik,Duhr:2011zq,Duhr:2012fh,2011arXiv1101.4497D}) to the amplitude evaluated
on a particular kinematic surface, after making a non-trivial identification between the two sets of kinematic
variables.

More specifically, the duality maps the three-point form factor evaluated at any
kinematic point to the six-point amplitude, or hexagonal Wilson loop, evaluated on the kinematic surface given by
\begin{equation}
  \label{eq:deltazero}
 \Delta(\hat u, \hat v, \hat w) \equiv
 (1-\hat u -\hat v - \hat w)^2 - 4 \hat u \hat v \hat w = 0\,.
\end{equation}
The variables $\hat u$, $\hat v$, $\hat w$ are the
dual-conformal cross ratios for the hexagon, defined by
\begin{equation}
  \hat u = \frac{s_{12}s_{45}}{s_{123}s_{345}} \,, \qquad
  \hat v = \frac{s_{23}s_{56}}{s_{234}s_{123}} \,, \qquad
  \hat w = \frac{s_{34}s_{61}}{s_{345}s_{234}}\,.
\end{equation}
On this kinematic slice, the momenta $\hat{p}_i$ of the six-point amplitude are related to each other by $\hat{p}_{i+3} = -\hat{p}_i$. Here and below, we use hats on the kinematic variables associated with the six-point amplitude in order to distinguish them from the kinematic variables associated with the form factor.

On the surface \eqref{eq:deltazero}, the three parity-odd letters in the hexagon symbol alphabet become unity,
\be
\hat{y}_u = \hat{y}_v = \hat{y}_w = 1,
\label{ygoto1}
\ee
and thus drop out of the symbol. A convenient basis in which to express the remaining letters is given by
\begin{align}
  \hat a
  &=
  \frac{\hat u}{\hat v \hat w}\,,&
  \hat b
  &=
  \frac{\hat v}{\hat w \hat u}\,,&
  \hat c
  &=
  \frac{\hat w}{\hat u \hat v}\,,\nonumber \\
  \hat d
  &=
  \frac{1-\hat u}{\hat u}\,,&
  \hat e
  &=
  \frac{1-\hat v}{\hat v}\,,&
  \hat f
  &=
  \frac{1-\hat w}{\hat w}\,.
\end{align}
In this basis, the kinematic map required to relate the form factor and the amplitude becomes
\begin{equation}
  \sqrt{\hat{a}}\ \Leftrightarrow\ d\,, \qquad\quad
  \hat{d}\ \Leftrightarrow\ a \,,
\label{eq:KinematicMap}  
\end{equation}
as well as its cyclic images. The precise statement of the duality is
\begin{equation}
\label{eq: statement of the duality}
S\left(\mathcal{E}_6 \big|_{\hat{a}\to d^2,\hat{d}\to a,\dots}\right)\ =\ \mathcal{E}_c\,,
\end{equation}
where $S$ is the antipode map.

Recursive algorithms exist for computing the antipode of generic polylogarithms, modulo factors proportional to $i \pi$; see for instance ref.~\cite{DelDuca:2016lad}. The relation \eqref{eq: statement of the duality} is thus understood to hold only up to such terms; see also the discussion in ref.~\cite{Dixon:2021tdw}. At the level of the symbol, the computation of the antipode becomes particularly simple -- one just reverses the
order of the letters in each symbol word, up to a sign:
\begin{equation}
 S(x_1\otimes\dots\otimes x_m)=(-1)^m x_m\otimes\dots\otimes x_1\,.
\end{equation}
In section \ref{sec:special_kinematics}, we study the extent to which the two quantities agree when the contributions proportional to $i\pi$ are also taken into account.

The antipodal duality allows the constraints discussed in section
\ref{sec:summary-constraints} to be additionally interpreted as properties of the
scattering amplitude, rather than just as conditions on the form factor. The
simplest of these properties are the extended Steinmann 
relations, which state that the letters $\hat{a}$
and $\hat{b}$ never appear in adjacent entries of the symbol,
as well as the dihedral images of this condition.
The kinematic map~(\ref{eq:KinematicMap})  translates this condition into the form factor adjacency condition~\eqref{eq:adjacency_restrictions_1}:
\begin{equation}
  \cancel{\dots \hat a \otimes \hat b \dots} \qquad
  \Leftrightarrow\qquad
  \cancel{\dots d \otimes e \dots} \,.
\label{eq:mapadj_1}
\end{equation}

The extended Steinmann relations, together with the branch-cut
conditions, also imply the non-adjacency of the letters
$\hat{a}$ and $\hat{d}$ in the six-point amplitude, plus
all dihedral images of these conditions.
The map~(\ref{eq:KinematicMap}) translates these additional conditions
into the form factor adjacency constraints~\eqref{eq:adjacency_restrictions_2}:
\begin{equation}
  \cancel{\dots \hat a \otimes \hat d \dots}  \quad , \quad \cancel{\dots \hat d \otimes \hat a \dots}  \qquad
  \Leftrightarrow\qquad
  \cancel{\dots d \otimes a \dots} \quad , \quad \cancel{\dots a \otimes d \dots} \, .
\label{eq:mapadj_2}
\end{equation}
Through the antipodal duality, the form factor adjacency relations can also be viewed as being implied by the cluster adjacency conditions associated with cluster coordinates on the Grassmannian $\operatorname{Gr}(4,6)$~\cite{Drummond:2017ssj}.

In section~\ref{sec:background}, it was further observed that the condition~\eqref{eq:triple_adjacency_restriction}
is obeyed by adjacent triplets of letters in the three-point form factor. This restriction goes beyond what is implied by the integrability condition of the form factor and the adjacency conditions~\eqref{eq:adjacency_restrictions_1} and~\eqref{eq:adjacency_restrictions_2}.  However, it turns out to be implied, via the antipodal duality, by the extended Steinmann relations and integrability conditions of the six-point amplitude. The appearance of a new triple adjacency constraint in the form factor space contrasts with the space of hexagon functions, where no extra restrictions appear beyond adjacent pair relations~\cite{Caron-Huot:2019bsq}.  Notice that there are three separate integrability conditions for functions of three variables, like hexagon functions, while there is only one integrability condition, $\partial_u \partial_v F(u,v) = \partial_v \partial_u F(u,v)$, for functions of two variables, like the form factor.  The additional integrability conditions lend more power to the pair relations for hexagon functions, such that they subsume the triple adjacency constraints in the dual form factor space.

We also observe a saturation of the allowed multiple final entries of
the form factor, as recorded by the blue entries in Table
\ref{tab:EEcopdimsymb}. As can be seen from Table \ref{tab:Cparameters},
the assumed relations on the final $(L-1)$ entries
work as powerful empirical constraints that keep the number of free
parameters of our ans\"atze under control; however, these constraints do
not seem to be well motivated by the physical principles underlying the
form factor. In the light of the duality, however, we can understand the
final $w$ entries of the form factor as
coinciding with the space of parity-even weight-$w$ hexagon symbols (after applying the map~\eqref{eq: statement of the duality}). These symbols
are highly constrained by (three-variable) integrability, the branch cut conditions \cite{Gaiotto:2011dt}, and the extended Steinmann/cluster adjacency conditions~\cite{Caron-Huot:2019bsq,Drummond:2017ssj}. In addition, the fact that the
OPE cannot result in irreducible multiple zeta values such as $\zeta_{5,3}$
at the points
$(\hat{u},\hat{v},\hat{w})=(1,0,0),(0,1,0),(0,0,1)$
starts to eliminate functions at weight eight~\cite{Caron-Huot:2019bsq}.  In fact, it eliminates precisely three
functions at this weight, taking the number of weight-eight parity-even hexagon
symbols from 282 down to 279, in striking agreement with the $\red{279}$
in Table~\ref{tab:EEcopdimsymb}.

While some properties of the form factor are currently better
understood on the amplitude side of the duality, it is also
possible to interpret the duality the other way around, and treat the
form factor as boundary data for the scattering amplitude. Through five 
loops, we find that there is a unique hexagon function in the three-dimensional
bulk $(\hat u, \hat v, \hat w)$ that is equal to the form factor on the
duality slice $\Delta = 0$.  This uniqueness through five loops does not
require any other assumptions, such as dihedral symmetry or final-entry
conditions. Starting from six loops, the knowledge of the form factor is not enough to fix the amplitude in full kinematics. Concretely, at six and at seven loops, there is precisely one dihedrally-symmetric hexagon function that vanishes on the parity-even subspace, called $Z$ and $\tilde{Z}$, respectively~\cite{Caron-Huot:2019vjl}. The parity-even first coproducts of these functions vanish; only the parity-odd first coproducts $Z^{\hat{y}_i}$ and $\tilde{Z}^{\hat{y}_i}$ are nonzero.  Using this fact, one can show that the functions actually vanish in the near-collinear limit until ${\cal O}(\hat{T}^2)$, so that OPE contributions with two flux-tube excitations are needed to determine them. The corresponding ambiguities can be resolved using for instance the behavior of the six-point amplitude at the origin $(\hat{u},\hat{v},\hat{w})=(0,0,0)$~\cite{Caron-Huot:2019vjl,Basso:2020xts}. If we relax dihedral symmetry, there is still only one function at six loops, but three functions at seven loops. It would be interesting to quantify the growth of this ambiguity at higher loop orders, and use the three-point form factor as a short-cut to the MHV six-point amplitude.


\section{Behavior in Kinematic Limits}
\label{sec:special_kinematics}

The ansatz for our bootstrap approach is most naturally formulated in the Euclidean region, where all four Mandelstam invariants are negative and the form factor is manifestly real. There is also a pseudo-Euclidean region where all Mandelstam invariants are positive. This region describes the decay into three gluons (for example) of a massive particle (analogous to the Higgs boson) that couples to the chiral part of the stress-tensor supermultiplet.  In the divergent BDS-like prefactor, there are terms in the Euclidean region that pick up phases as they are continued into the pseudo-Euclidean region; however, the finite quantities $\mathcal{E}$ and $R$ remain real. On the other hand, we can analytically continue our result for the form factor into kinematic regions that describe two-to-two scattering processes. 

\begin{figure}[t]
\begin{center}
\includegraphics[width=5.5in]{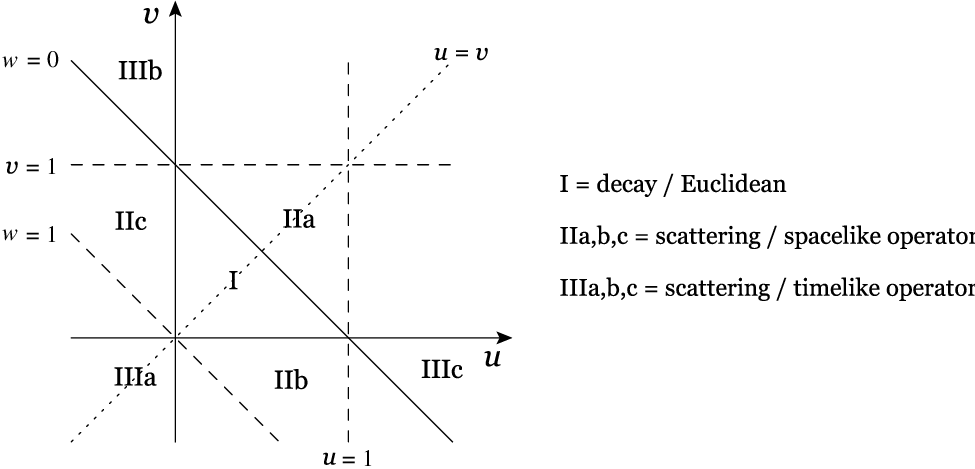} 
\end{center}
\caption{The two-dimensional space of kinematics on which the three-point form factor depends, in the $u$ and $v$ variables. The Euclidean region, where all Mandelstams are negative, is given by the interior of the triangle defined by $0 < u < 1$ and $0 <  v < 1-u$. The solid edges bounding the Euclidean region correspond to the collinear limits, near which FFOPE data can be accessed. We describe the other physical regions in the text.}
\label{Fig:uvplane}
\end{figure}

The full space of kinematics corresponding to real values of $u$ and $v$ is depicted in Figure~\ref{Fig:uvplane}. In this figure, the Euclidean region, which satisfies $0 < u,v,w < 1$, is labelled as region I. The three adjacent regions, denoted by IIa, IIb, and IIc, describe scattering processes in which the momentum of the operator is space-like, mimicking deep inelastic scattering. Region IIa is delimited by $w < 0 < u, v$, while regions IIb and IIc satisfy cyclic permutations of these inequalities. Finally, regions IIIa, IIIb, and IIIc describe two-to-two scattering processes in which a time-like operator is produced. Region IIIa is delineated by $u, v < 0 < w$, and IIIb and IIIc are defined by cyclic sets of conditions.  The solid lines in the figure are the collinear limits, but they are also the location of branch cuts, where $\mathcal{E}$ acquires imaginary parts as one leaves region I.

While in generic kinematics, $\mathcal{E}$ is a complicated function of two arguments, it simplifies dramatically in special kinematic limits. Of particular interest are kinematic lines along which it simplifies to HPLs~\cite{Remiddi:1999ew}, since these functions are especially simple to work with, and the number of them grows more slowly than 2dHPLs. Still, the number of HPLs with indices $\{0,1\}$ required at $L$ loops is $2^{2L-2}$ at symbol level, because the symbol is a binary string of length $2L$, and the ``$-2$'' is associated with initial- and final-entry conditions.  At eight loops, there are $16384$ possible terms in the answer (and almost all of them have nonzero coefficients).  For HPLs with indices $\{-1,0,1\}$, the number of HPLs grows even more rapidly, leading to tens of millions of terms.  For this reason, some care needs to be taken in their computation, and their series expansion for rapid enough evaluation.

We are also interested in the number-theoretic content of $\mathcal{E}$ at kinematic points where it can be expressed in terms of well-studied transcendental constants such as MZVs or alternating sums. These are the easiest points at which to search for a non-trivial coaction principle, such as has been seen in similar contexts~\cite{Schlotterer:2012ny,Schnetz:2013hqa,Brown:2015fyf,Panzer:2016snt,Schnetz:2017bko,Caron-Huot:2019bsq,Gurdogan:2020ppd}.  For this purpose, it is necessary to reduce a large number of possible sums, or iterated integrals, into a basis, and we have used the {\sc Maple} package {\sc HyperlogProcedures}~\cite{HyperlogProcedures} for this purpose.  We also need this functionality at points that serve as boundary conditions, as we will discuss shortly.


\subsection{Special Lines}
\label{subsec:lines}

\begin{figure}[t]
\begin{center}
\includegraphics[width=3in]{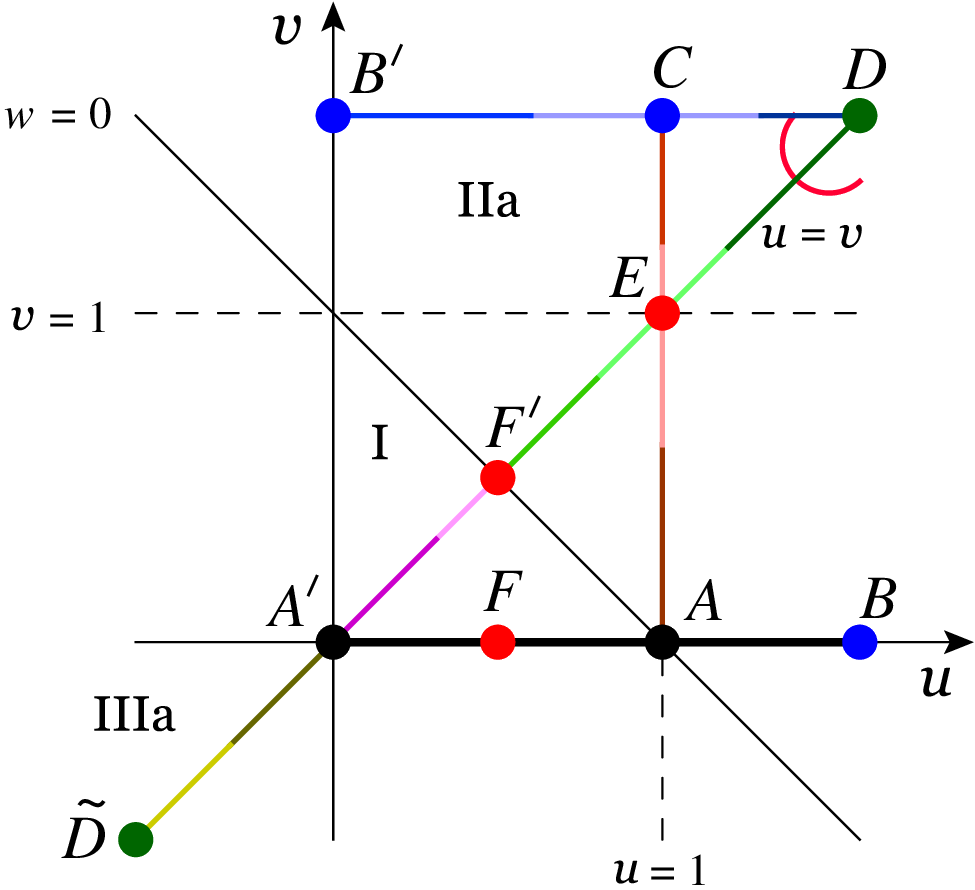} 
\end{center}
\caption{Various lines in the $(u,v)$ plane along which the form factor can be expressed in terms of HPLs. On the black and blue lines, the HPLs have indices $\{0,1\}$; on the other lines the indices are $\{-1,0,1\}$. Values at the black and blue points are all MZVs. At the red points they are alternating sums, and at the green points they could be either, depending on the direction of approach. See the text for further details.}
\label{Fig:uvplanelines}
\end{figure}

We begin by exploring the behavior of our new results on several lines, along which the form factor can be expressed in terms of HPLs. First, we consider the two $(u,v,w)$ lines parametrized by $(u,u,1-2u)$ and $(1,v,-v)$, which were studied through five loops in ref.~\cite{Dixon:2020bbt}.  Then we explore the behavior of $\mathcal{E}$ along a third HPL line, which is parametrized by $u$ in the $v \to \infty$ limit. These lines of interest are depicted in the $(u,v)$ plane in Figure~\ref{Fig:uvplanelines}.

This figure also indicates roughly how $\mathcal{E}$ was constructed on these lines. Because $\mathcal{E}$ was constrained in the strict-collinear and near-collinear limit, the first place it is known, at any loop order, is on the $u$-axis, shown in heavy black, particularly the segment between $A$ and $A'$ where it is real. The point $A$ is the reference base point $(u,v) = (1,0)$ where all the constant values of the basis functions are defined.  The point $A'$ is related to $A$ by a dihedral symmetry, but they are also connected by integrating along the line that interpolates between them. This gives rise to dihedral constraints on the constants that appear at these points, which involve MZVs as the form factor evaluates to HPLs with indices $\{0,1\}$ on this line, which get evaluated at unity at the point $A'$.  (There are also singular logarithms, powers of $\ln v$, to track along this line.)  The next step is to analytically continue $u$ past 1. To do this, one uses a suitable HPL representation where the only discontinuous function is $\ln w \approx \ln(1-u) \to \ln(u-1) - i\pi$.  Then one integrates up the continued function from $u=1^+$ to
$u=\infty$, which is point $B$. All the constants at $B$ are MZVs.  One can systematically evaluate all the constants at $B$ through weight 8.  Beyond weight 8, where we do not have a function-level basis, we evaluate the constants only for the independent coproducts of $\mathcal{E}^{(L)}$.  For $\mathcal{E}^{(8)}$, that includes 155 independent heptuples, 85 independent hextuples, 45 independent quintuples, and so on.  Such higher-weight constants have to be determined at all of the labelled points in Figure~\ref{Fig:uvplanelines}.

Dihedral symmetry maps the function values at $B$ to those at $B'$.
Point $B'$, at $u=0$ and $v=\infty$ is the boundary condition for integrating up along the region II line $v=\infty$ at fixed $u$.  It is easy to see that the alphabet $\cL_u$ collapses to $\{u,1-u\}$ on this line, or equivalently $\{x,1-x\}$ with $x \equiv 1-1/u$. Using $x$ as a variable exploits both the initial- and final-entry conditions, so the HPLs indices on this line always start with 0 and end with 1.  The first few orders are given by
\bea
\mathcal{E}^{(1)}(v\to\infty) &=& 2 H_{0,1} + 4 \zeta_2 \,, \label{eq:E1RII}\\
\mathcal{E}^{(2)}(v\to\infty) &=& -8 H_{0,0,0,1}-4 H_{0,1,1,1}
+ 8 \zeta_2 H_{0,1} - 3 \zeta_4 \, \label{eq:E2RII}\\
\mathcal{E}^{(3)}(v\to\infty) &=&
96 H_{0,0,0,0,0,1}+16 H_{0,0,0,1,0,1}+16 H_{0,0,0,1,1,1}
+16 H_{0,0,1,0,0,1}+8 H_{0,0,1,0,1,1}\nonumber\\ &&\null 
+8 H_{0,0,1,1,0,1}+16 H_{0,1,0,0,0,1}+8 H_{0,1,0,0,1,1}+12 H_{0,1,0,1,0,1}
+4 H_{0,1,0,1,1,1}\nonumber\\ &&\null 
+8 H_{0,1,1,0,0,1}+4 H_{0,1,1,0,1,1}+4 H_{0,1,1,1,0,1}
+24 H_{0,1,1,1,1,1}\nonumber\\ &&\null 
-\zeta_2 (16 H_{0,0,0,1}+8 H_{0,0,1,1}+4 H_{0,1,0,1}+44 H_{0,1,1,1})
\nonumber\\ &&\null 
-\zeta_3 (8 H_{0,0,1}-4 H_{0,1,1})
-85 \zeta_4 H_{0,1}
-\frac{455}{12} \zeta_6 + 10 (\zeta_3)^2 \,,\label{eq:E3RII}
\eea
where the omitted HPL argument is $x$.
This line is antipodally dual to the hexagon line $(\hat{u},\hat{v},\hat{w})=(1,\hat{v},\hat{v})$ with $1/\hat{v} = x = 1-1/u$~\cite{Dixon:2021tdw}.
(In fact, this line is where the duality was first spotted.)
The first line of Table~\ref{tab:HPL_counts} shows the number of HPLs
as a function of loop order $L$ at symbol level,
i.e.~setting all the zeta values to zero.

The HPL representation allows determination of the constants at points $C$ and $D$, which are all MZVs.  $D$ is the $u\to\infty$ limit of the $v\to\infty$ line.
To plot a high-loop-order function on this line, the evaluation of so many high-weight HPLs becomes impractical, and it is best to perform three series expansions around $u=0,1,\infty$ ($B'$, $C$, $D$), as indicated by the three different shades of blue on this line.  The three series expansions have overlapping radii of convergence, so high numerical accuracy can be achieved with a modest number of terms.\footnote{Around 60 terms give 10--15 significant digits, depending on the loop order and the overlap region.}

\begin{table}[t]
\begin{center}
\begin{tabular}{c||c|c|c|c|c|c|c|c}
$L$ & 1 & 2 & 3 & 4 & 5 & 6 & 7 & 8 \\
\hline\hline
$(u,\infty,-\infty)$ HPLs & 1 &  2 & 14 & 61 & 252 & 1,018 & 4,092 & 16,380 \\
\hline
$(1,v,-v)$ HPLs & 2 &  4 &  54 &   318 &  1,946 &  11,432 & 66,912 & 390,046 \\
\hline
$(u,u,1-2u)$ HPLs & 4 & 25 & 269 & 2,580 & 24,484 & 221,249 & 1,992,784 & 17,936,054 
\end{tabular}
\caption{\label{tab:HPL_counts} The number of distinct HPLs at symbol level that appear in the expression for $\mathcal{E}$ on three lines:
$(u,\infty,-\infty)$, $(1,v,-v)$, and $(u,u,1-2u)$. The number depends some on which HPL argument is used; here we use $1-1/u$, $1/v$, and $u$, respectively.}
\end{center}
\end{table}

The line $u=1$ is more challenging than $v\to\infty$ because its HPL indices are $\{-1,0,1\}$, leading to a larger growth in the number of terms, as shown in the second line of Table~\ref{tab:HPL_counts}.  While the function values at points $C$ and $A$ are MZVs, the values at $E$ are alternating sums. One can obtain the values at $E$ by integrating up from $A$, or down from $C$, providing a nice consistency test.  Again three separate series expansions around $v=0,1,\infty$ ($A$, $E$, $C$) have overlapping ranges of convergence.

The symmetric line $u=v$ also has HPL indices $\{-1,0,1\}$, but typically many more terms are encountered than for the line $u=1$, as shown in the third line of Table~\ref{tab:HPL_counts}.  Also, there are three separate regions to cover, I, IIa, and IIIa, and a total of seven different series expansions are required, as indicated by the shadings in the figure.  The point $F'$ is related to $F$ by dihedral symmetry. Values at $F$ are obtained by evaluating HPLs with indices $\{0,1\}$ at $u=1/2$, which gives rise to alternating sums.

Two of the points we need to expand around are $D$ and $\tilde{D}$.  Due to its final-entry condition, the form factor has the same value at these two points; however, in general the derivatives (coproducts) of the form factor do not have this property, so these values have to be determined separately.  Similarly, the final-entry condition means that the value of $\mathcal{E}$ at $D$ does not depend on the direction of approach, but the same is not true for the coproducts of the form factor.  The red arc near point $D$ is meant to depict an auxiliary line corresponding to sending $u,v\to\infty$ while holding their ratio $r=u/v$ fixed.  In order to fix the necessary constants at infinity along the symmetric line $u=v$, one can compute the dependence of all functions on $r$, and then integrate from $r=0$ (the point $D$ as approached along the $v=\infty$ line) to $r=1$ (the point $D$ as approached along the symmetric line).

There are so many high-weight HPLs on the symmetric line that an alternative strategy for generating their series expansion is necessary.  The coproduct relations encode the multiple derivatives of the functions everywhere, including at the points around which one wishes to series expand.  It is not necessary to have the full HPL representation of the function in order to series expand it.  It is sufficient to have the representation of enough of its multiple coproducts, for the line along which one wishes to expand.  In practice we computed the quintuple coproducts, which at eight loops are weight-11 HPLs, whose series expansion of order 60 terms could still be obtained quickly enough.  Then we integrated up the relevant series term by term, five times in a row.

\begin{figure}[t]
\centering
 \begin{subfigure}[b]{0.45\textwidth}
         \centering
         \rotatebox{90}{\tiny \hspace{0.12\paperwidth}\clap{$\frac{R^{(L)}(u,u,1-2u)}{R^{(L)}\left(\frac13,\frac13,\frac13\right)}$}}
         \includegraphics[height=0.23\paperwidth,trim=3.97cm 0.24cm 2.3cm 0.28cm,clip]{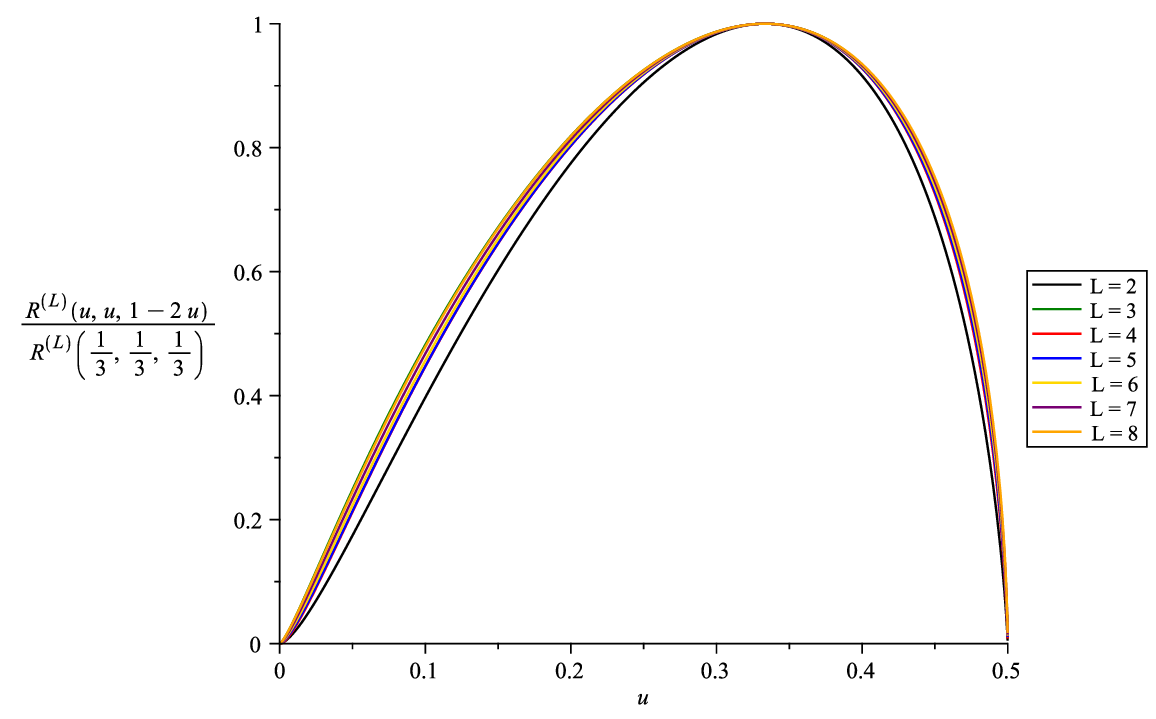}
         \caption{\phantom{.}}
         \label{Fig:RLsym}
     \end{subfigure}
     \hfill
      \begin{subfigure}[b]{0.54\textwidth}
         \centering
         \rotatebox{90}{\tiny \hspace{0.12\paperwidth}\clap{$\frac{{\mathcal{E}}^{(L)}(u,u,1-2u)}{{\mathcal{E}}^{(L-1)}(u,u,1-2u)}$}}
         \includegraphics[height=0.23\paperwidth,trim=4.72cm 0.2cm 0.1cm 0.34cm,clip]{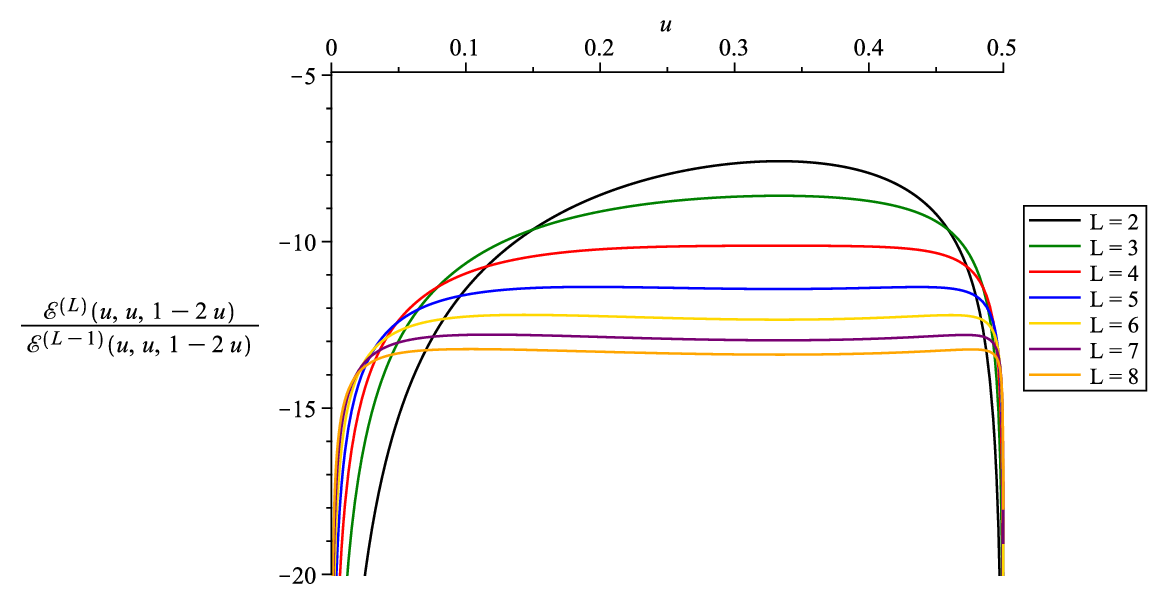}
         \caption{\phantom{.}}
         \label{Fig:ELsym}
     \end{subfigure}
\caption{(a) The remainder function $R^{(L)}$ along the $u=v$ line within the Euclidean region, which corresponds to $0 < u < 1/2$. We have normalized $R^{(L)}$ at each loop order by the maximum value it takes along this segment, which occurs when $u = v= w = 1/3$. (b) The ratio of $\mathcal{E}$ at successive loop orders along the $u=v$ line, within the Euclidean region.}
\label{Fig:RLsym8loops}
\end{figure}

\begin{figure}[t]
\centering
      \begin{subfigure}[b]{0.45\textwidth}
         \centering
         \rotatebox{90}{\tiny \hspace{0.12\paperwidth}\clap{$\frac{R_6^{(L)}(u)}{ R^{(L)}(u)}$}}
         \includegraphics[height=0.27\paperwidth,trim=2.2cm 0.2cm 4.1cm 0.3cm,clip]{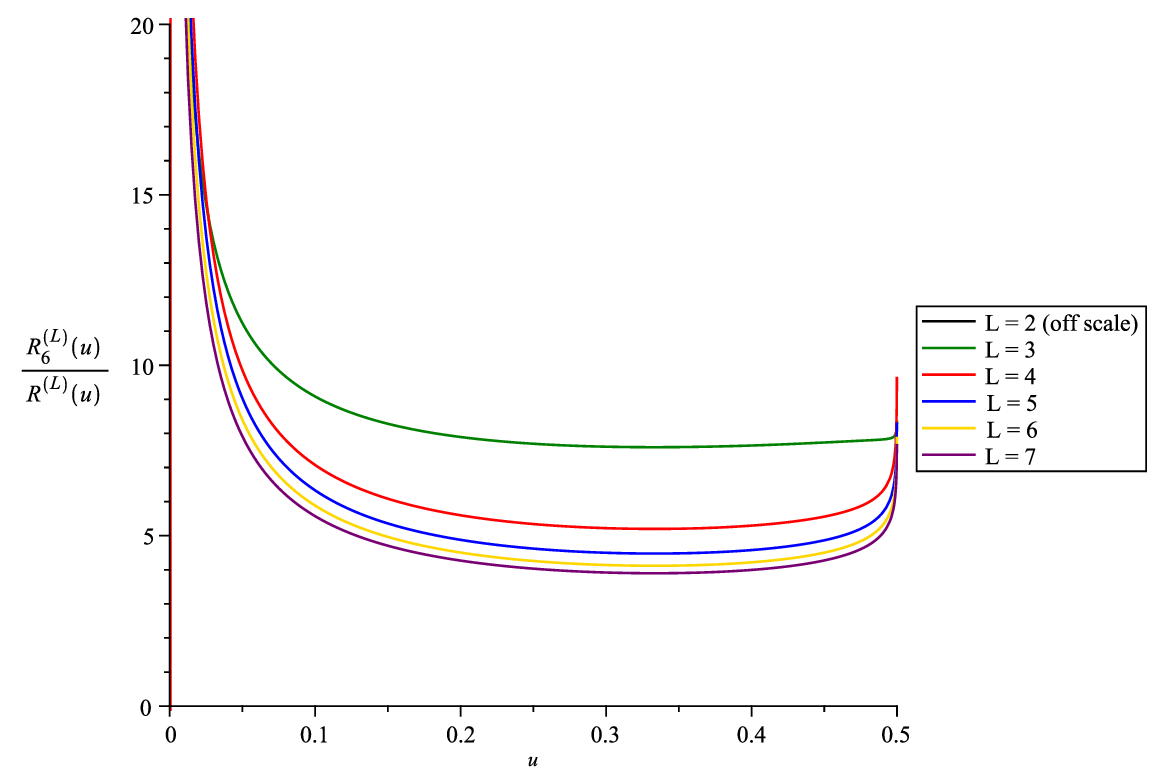}
         \caption{\phantom{.}}
         \label{Fig:RratsymI}
     \end{subfigure}
 \begin{subfigure}[b]{0.54\textwidth}
         \centering
         \rotatebox{90}{\tiny \hspace{0.12\paperwidth}\clap{$\frac{{\cal E}_6^{(L)}(u)}{{\cal E}_c^{(L)}(u)}$}}
         \includegraphics[height=0.27\paperwidth,trim=2.4cm 0.2cm .1cm 0.3cm,clip]{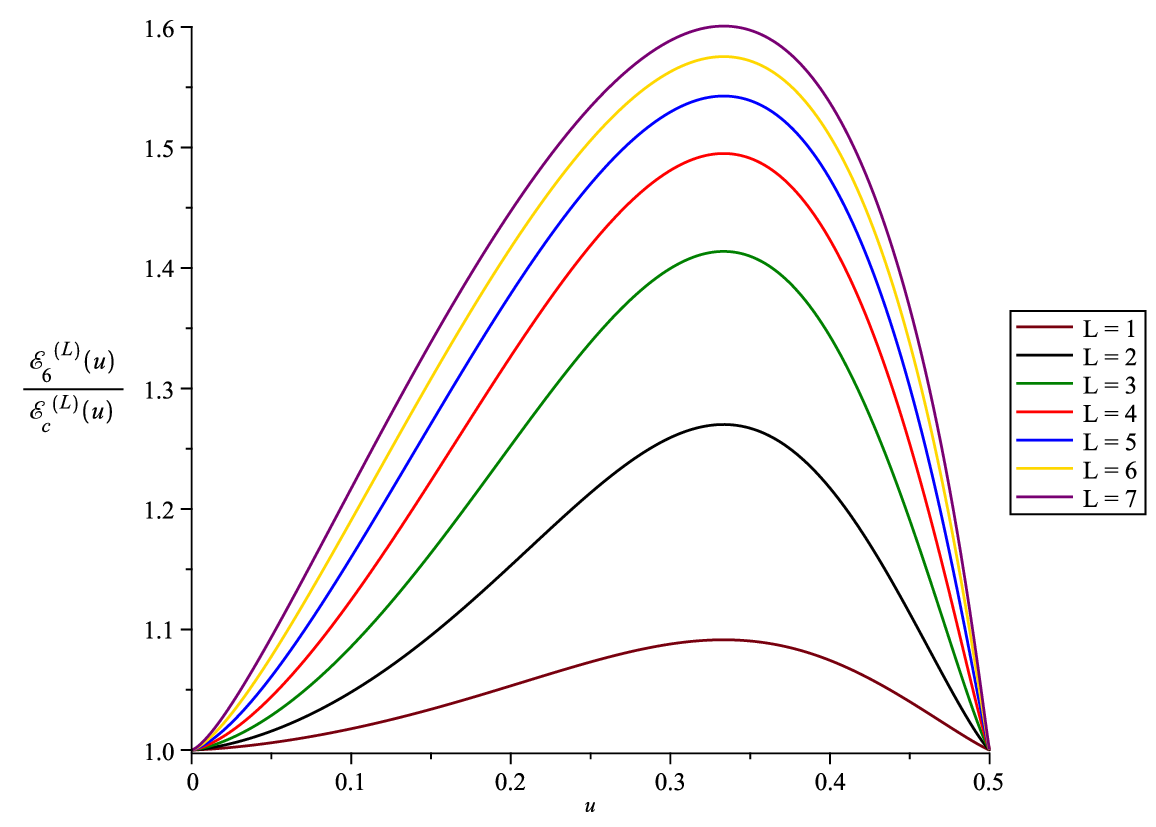}
         \caption{\phantom{.}}
         \label{Fig:EratsymI}
     \end{subfigure}
\caption{The ratio of the six-point MHV amplitude and the three-point form factor, evaluated on different sides of the antipodal duality on the line $(u,u,1-2u)$ in the form factor variables. The six-point remainder function is denoted by $R_6^{(L)}(u)$, while the cosmically and BDS-like normalized amplitude is denoted by ${\cal E}_6^{(L)}(u)$.}
\label{Fig:duality_numerical_comparison}
\end{figure}

Consider first the $u=v$ line, which passes through regions I, IIa, and IIIa. On this line, the remainder function can be expressed in terms of HPLs of argument $2u - 1$ with indices drawn from $\{-1,0,1\}$. It must vanish at $u = 0$ and $u=1/2$, since at these points the line intersects a collinear limit. Between these values, the remainder function is manifestly real.  We plot it through eight loops in Figure~\ref{Fig:RLsym}. In this plot, $R^{(L)}$ has been normalized by its maximum value along this line segment, which must occur at $u=v=w=1/3$ due to symmetry. We see that the overall shape of $R^{(L)}$ stays remarkably stable as one goes up in loop order. In Figure~\ref{Fig:ELsym}, we plot the ratio of the $\mathcal{E}$ at successive loop orders, $\mathcal{E}^{(L)}/\mathcal{E}^{(L-1)}$, along the same line segment. The ratio exhibits a remarkable constancy in $u$, over a broader and broader range of $u$ as $L$ increases.  The ratio is singular at both ends, because $\EE^{(L)}$ diverges as the $2L^{\rm th}$ power of a logarithm; see \eqn{eq:strictcoll}.  However, in the bulk we see the expected finite radius of convergence of planar $\mathcal{N}=4$ sYM theory taking over.
While the ratio between loop orders continues to decrease, the ratio of eight to seven loops is still far from the value $-16$, the expected asymptotic ratio for many quantities in planar $\mathcal{N}=4$ sYM theory.  Yet we will see in the next subsection that it behaves similarly at $u=1/3$ to the cusp anomalous dimension.

We can also compare the form factor on the $u=v$ line to the numerical value of the six-point MHV amplitude on the other side of the antipodal duality~\cite{Dixon:2021tdw}.  The kinematic map sends the form factor line $(u,u,1-2u)$ to the six-point amplitude line $(\hat{u},\hat{u},(1-2\hat{u})^2)$, where $\hat{u}=(1-2u)/2/(1-u)$.  For general kinematics, the duality is only known to hold up to factors proportional to $i \pi$. Even if these factors were included, one would not expect these quantities to exhibit matching behavior, due to the antipodal switch of discontinuities and derivatives, or in other words, the reversal of the order of integration.  However, we observe in Figure~\ref{Fig:duality_numerical_comparison} that the ratio of these quantities behave in a regular way as one moves up in loop order. Figure~\ref{Fig:RratsymI} plots the remainder functions $R_6$ and $R$ for the six-point amplitude and the three-point form factor, respectively. Figure~\ref{Fig:EratsymI} plots $\mathcal{E}_6^{(L)}$ for the amplitude and the cosmic normalization $\mathcal{E}_c^{(L)}$ of the form factor.  This normalization matches exactly at the soft-collinear endpoints at one loop, and the relation~\eqref{eq:calE_def} to the remainder function(s), which vanish at the endpoints, ensures that $\mathcal{E}_6^{(L)}$ equals $\mathcal{E}_c^{(L)}$ at the two endpoints at any loop order, as seen in Figure~\ref{Fig:EratsymI}.  The endpoint behavior in Figure~\ref{Fig:RratsymI} is more singular because numerator and denominator both vanish, with different subleading-power behavior. The more remarkable property of Figure~\ref{Fig:EratsymI} is that the ratio never gets larger than $1.6$, even though the individual components are changing by over a factor of 10 per loop (Figure~\ref{Fig:ELsym}).

\begin{figure}[t]
\centering
 \begin{subfigure}[b]{0.45\textwidth}
         \centering
         \rotatebox{90}{\tiny \hspace{0.12\paperwidth}\clap{$\frac{\text{Re } R^{(L)}(u,u,1-2u)}{\text{Re } R^{(L)}(1,1,-1)}$}}
         \includegraphics[height=0.23\paperwidth,trim=4.9cm 0.2cm 2.38cm 0.1cm,clip]{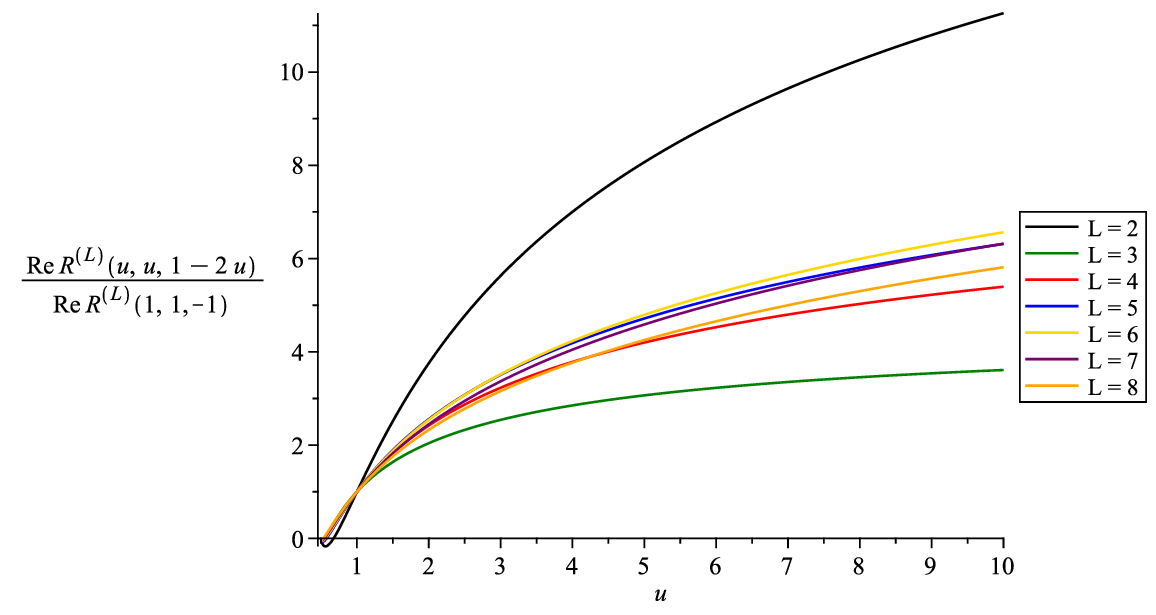}
         \caption{\phantom{.}}
         \label{Fig:ReRLsymII}
     \end{subfigure}
     \hfill
      \begin{subfigure}[b]{0.54\textwidth}
         \centering
         \rotatebox{90}{\tiny \hspace{0.12\paperwidth}\clap{$\frac{\text{Im } R^{(L)}(u,u,1-2u)}{\text{Re } R^{(L)}(1,1,-1)}$}}
         \includegraphics[height=0.23\paperwidth,trim=4.9cm 0.2cm 0.1cm 0.35cm,clip]{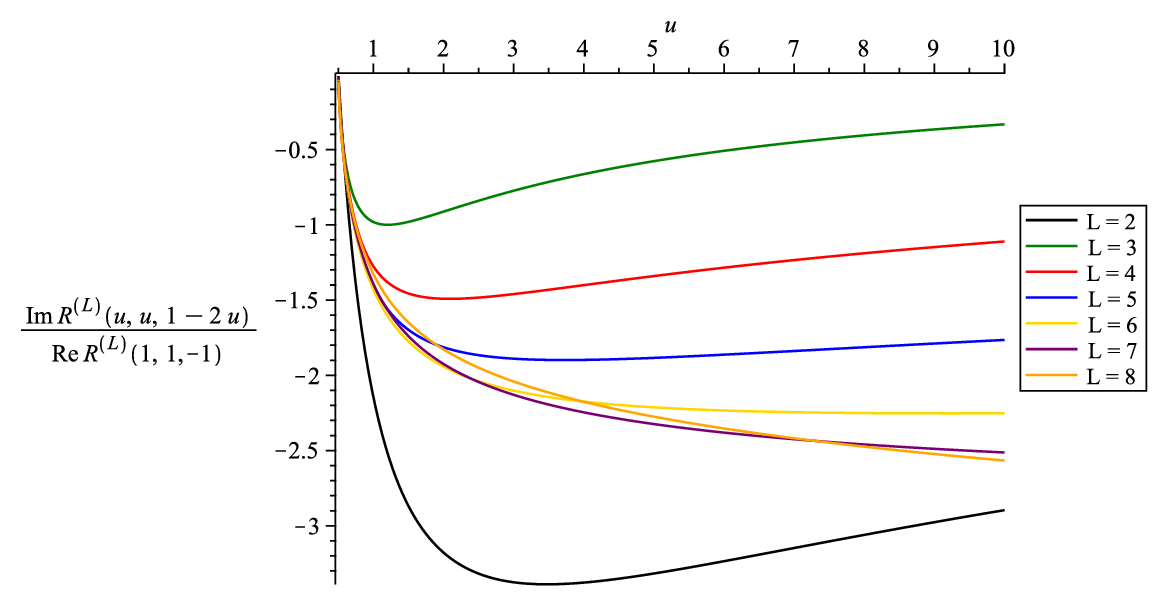}
         \caption{\phantom{.}}
         \label{Fig:ImRLsymII}
     \end{subfigure}
\caption{The real (a) and imaginary (b) parts of the remainder function $R^{(L)}$ on the $u=v$ line in scattering region IIa, where $u>1/2$. We have normalized these functions by the value of their real part at $u=1$.}
\label{Fig:RLsymII}
\end{figure}

To obtain $R^{(L)}$ along the $u=v$ line in region IIa, where $u>1/2$, we analytically continue out of the Euclidean region by sending $\ln w \to \ln |w| - i \pi$ near the $w=0$ line. Starting from the coproduct representation of $R^{(L)}$ (or actually $\mathcal{E}^{(L)}$), we can then integrate up each weight iteratively to get the full functional form along this line. The integration constant required for each of these integrations at point $F'$ is determined by integrating all the functions in our basis along the collinear $w = 0$ line from $u = 1$ (where the integration constants are first defined) to $u = 1/2$. We plot the real and imaginary parts of the remainder function for $u > 1/2$ in Figure~\ref{Fig:RLsymII}.  The large negative imaginary part at two loops in Figure~\ref{Fig:ImRLsymII} is due to the normalizing denominator ${\rm Re}\,R^{(L)}(1,1,-1)$ being accidentally small at that loop order. (It has a zero crossing very near $u=1$.)  Otherwise, the behavior is rather regular in $L$, beyond two or three loops.

\begin{figure}[t]
\centering
 \begin{subfigure}[b]{0.45\textwidth}
         \centering
         \rotatebox{90}{\tiny \hspace{0.12\paperwidth}\clap{$\frac{\text{Re } R^{(L)}(-u,-u,1+2u)}{R^{(L)}(-u,-u,1+2u)|_{u \to \infty}}$}}
         \includegraphics[height=0.23\paperwidth,trim=5.3cm 0.1cm 2.2cm 0.1cm,clip]{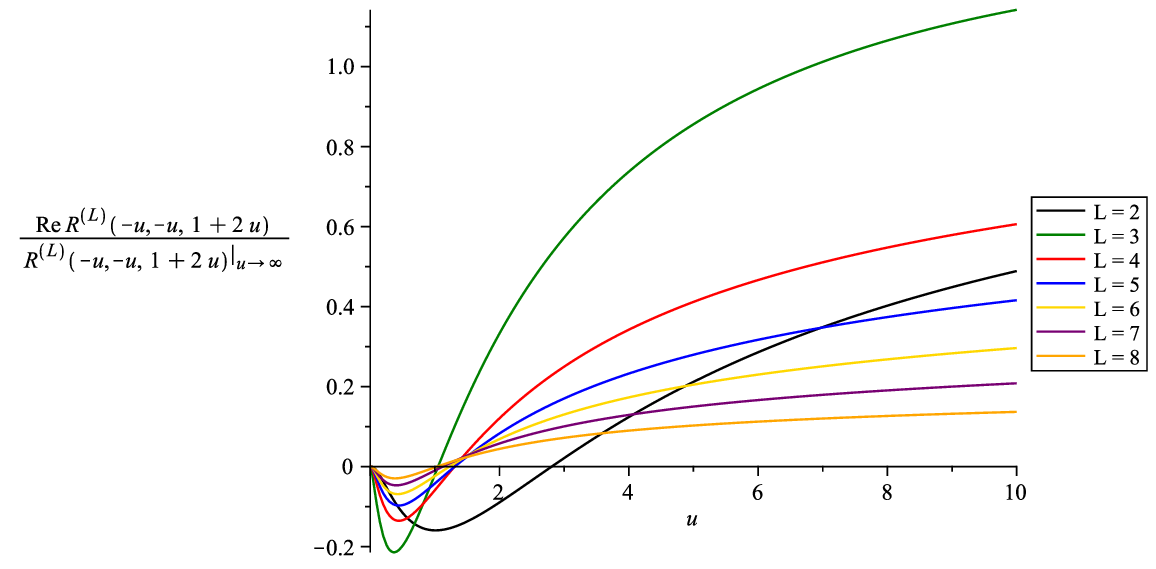}
         \caption{\phantom{.}}
         \label{Fig:ReRLIII}
     \end{subfigure}
     \hfill
      \begin{subfigure}[b]{0.54\textwidth}
         \centering
         \rotatebox{90}{\tiny \hspace{0.12\paperwidth}\clap{$\frac{\text{Im } R^{(L)}(-u,-u,1+2u)}{R^{(L)}(-u,-u,1+2u)|_{u \to \infty}}$}}
         \includegraphics[height=0.23\paperwidth,trim=5.3cm 0.1cm 0.1cm 0.2cm,clip]{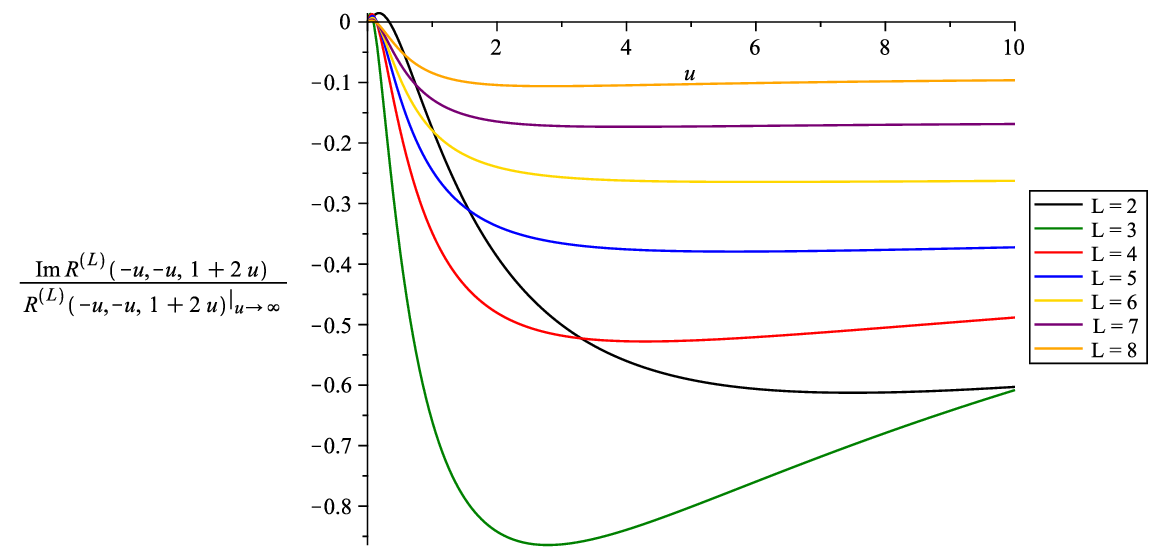}
         \caption{\phantom{.}}
         \label{Fig:ImRLIII}
     \end{subfigure}
\caption{The real (a) and imaginary (b) parts of the remainder function $R^{(L)}$ on 
  the $u=v$ line in region IIIa. We have normalized these functions by their value at $u \to \infty$, which is real.}
\label{Fig:RLIII}
\end{figure}

A similar procedure allows us to compute $R^{(L)}$ for $u = v < 0$. To do this, we analytically continue $\ln u \to \ln |u| - i \pi$ and $\ln v \to \ln |v| - i \pi$, and integrate up the coproduct representation using boundary conditions at the point $A'$, which are related by dihedral symmetry to the values at the base point $A$.  The real and imaginary parts of the remainder function are plotted in this region in Figure~\ref{Fig:RLIII}, normalized by the (real) value at infinity.  Again the behavior is rather regular at high loop order.  The asymptotic $u\to\infty$ behavior is not achieved until far beyond the plot range.

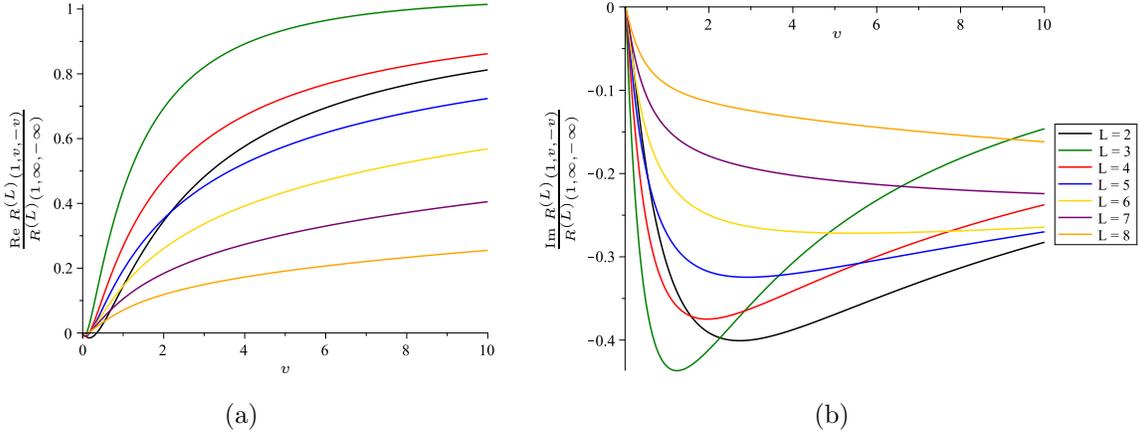
\begin{figure}[t]
\centering
 \begin{subfigure}[b]{0.45\textwidth}
         \centering
         \rotatebox{90}{\tiny \hspace{0.12\paperwidth}\clap{$\frac{\text{Re } R^{(L)}(1,v,-v)}{ R^{(L)}(1,\infty,-\infty)}$}}
         \includegraphics[height=0.23\paperwidth,trim=4.1cm 0.2cm 2.65cm 0.2cm,clip]{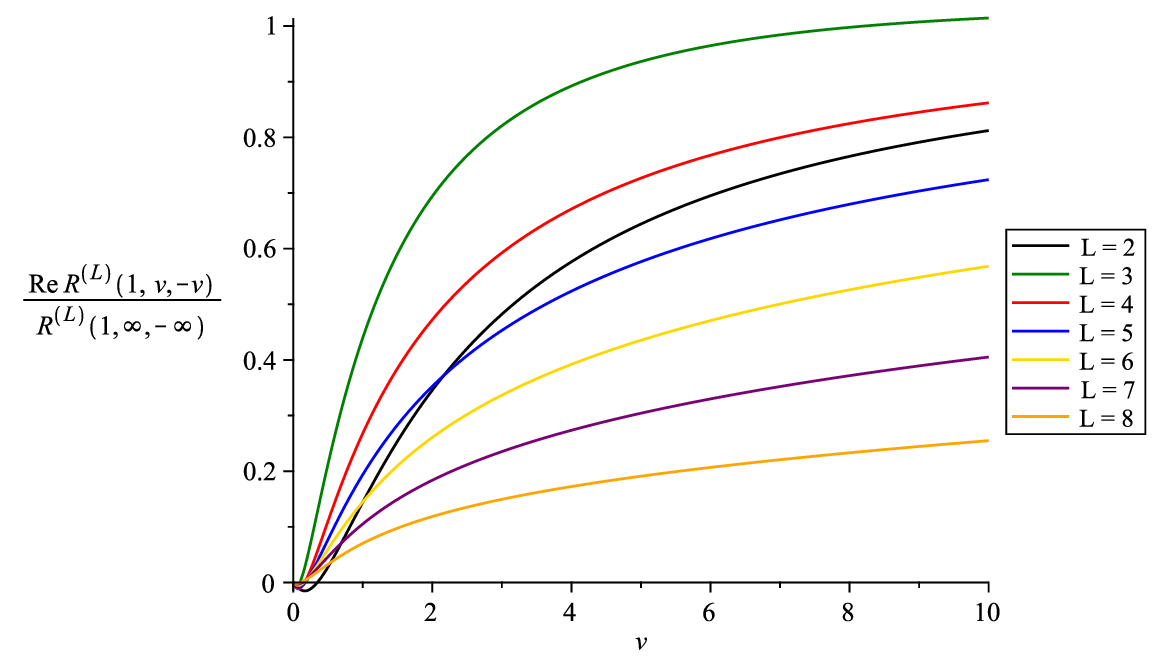}\raisebox{-0.7ex}{\smash{\llap{\begin{tikzpicture}\node[] at (0,10) {};  
              \node[fill=white] at (-2.65,-1) {$\scriptscriptstyle v$};                                                                                              \end{tikzpicture}%
}}}
         \caption{\phantom{.}}
         \label{Fig:ReRLII}
     \end{subfigure}
     \hfill
      \begin{subfigure}[b]{0.54\textwidth}
         \centering
         \rotatebox{90}{\tiny \hspace{0.12\paperwidth}\clap{$\frac{\text{Im } R^{(L)}(1,v,-v)}{ R^{(L)}(1,\infty,-\infty)}$}}
         \includegraphics[height=0.23\paperwidth,trim=4.1cm 0.1cm 0.1cm 0.25cm,clip]{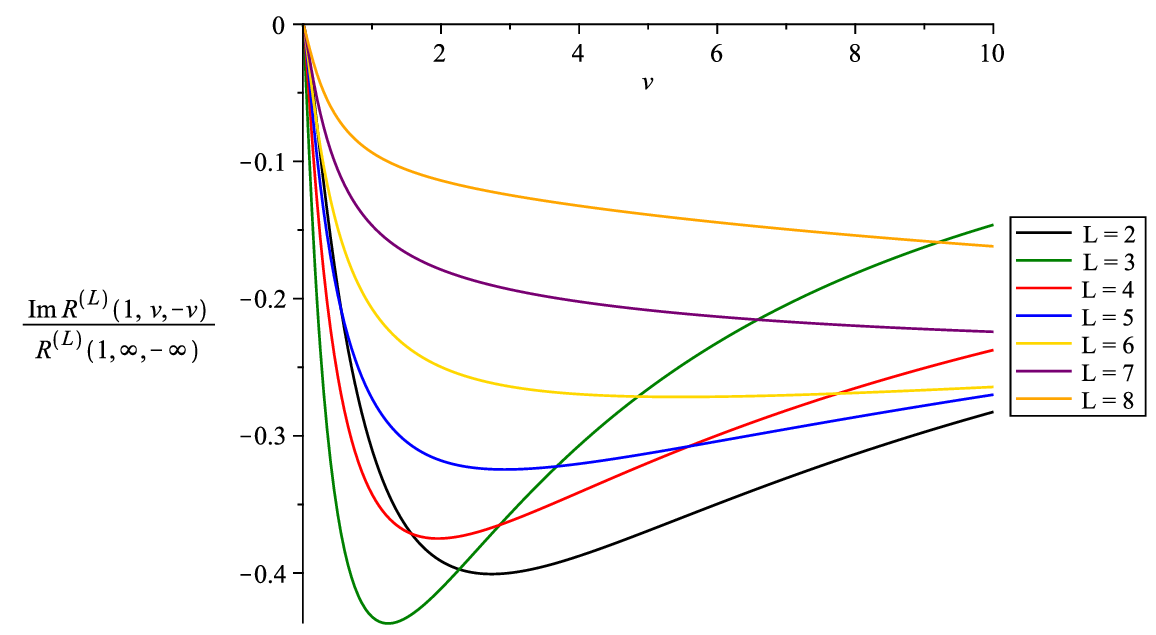}\raisebox{-0.7ex}{\smash{\llap{\begin{tikzpicture}\node[] at (0,0) {};  
              \node[fill=white] at (-3.9,4.5) {$\scriptscriptstyle v$};                                                                                              \end{tikzpicture}%
}}}
         \caption{\phantom{.}}
         \label{Fig:ImRLII}
     \end{subfigure}
\caption{The real (a) and imaginary (b) parts of the remainder function $R^{(L)}$ on the line $u=1$ in scattering region IIa, where $v>0$. We normalize these functions
  by their value at $v=\infty$, where they are real.
  }
\label{Fig:RLII}
\end{figure}

Next, we consider the $u=1$ line within region IIa, on which the remainder function can be expressed in terms of HPLs of argument $v$ with indices drawn from $\{-1,0,1\}$. We obtain the value of the remainder function on this line using the same analytic continuation as used to get into region IIa above, and then integrating up from the $v = 0$ point $A$, where all integration constants are defined.  We also integrate down from point $C$ at $v=\infty$, and get the same value at $v=1$ (point $E$) either way. We plot the real and imaginary parts of $R^{(L)}$ along this line in Figure~\ref{Fig:RLII}.  Again we normalize by the real value at $v=\infty$.  Again the imaginary parts are negative relative to this normalization, and the asymptotic behavior as $v\to\infty$ is not yet visible in the plot.

\begin{figure}[t]
         \centering
         \rotatebox{90}{\tiny \hspace{0.23\paperwidth}\clap{\large $\frac{R^{(L)}(u,v\to \infty, w\to - \infty)}{ R^{(L)}(u \to \infty ,v\to \infty,w\to - \infty)}$}}
         \includegraphics[width=5in,trim=4.2cm 0.1cm 0.1cm 0cm,clip]{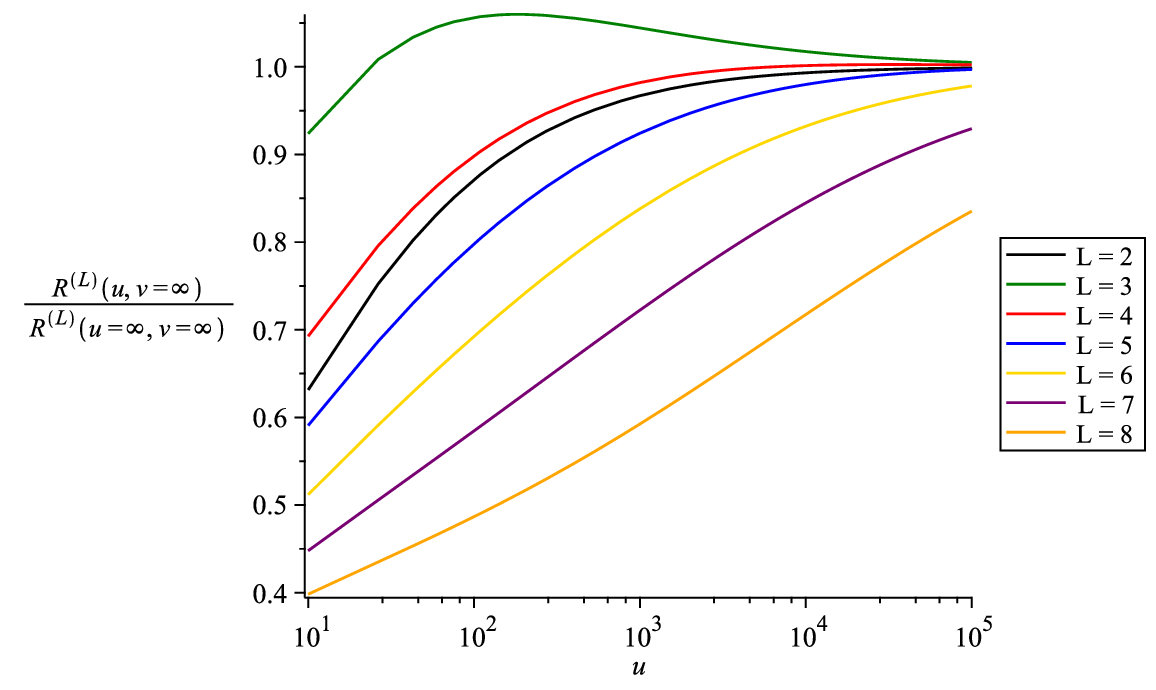}
\caption{The remainder function $R^{(L)}$ in region IIa in the limit $v \to \infty$ at fixed $u$, from two to eight loops. We have normalized $R^{(L)}$ at each loop order by its value at $u \to \infty$.  }
\label{Fig:RLregge8loops}
\end{figure}

Finally, let us consider the $v \to \infty$ limit of the remainder function at fixed $u$. $R^{(L)}(u)$ is manifestly real in this limit, and can be expressed in terms of HPLs of argument $u$ that involve indices $\{0,1\}$, as seen in eqs.~\eqref{eq:E1RII}--\eqref{eq:E3RII} for $\mathcal{E}^{(1)}$ to $\mathcal{E}^{(3)}$. We plot the value of the remainder function on this line for a large range of values of $u$ in Figure~\ref{Fig:RLregge8loops}. We normalize each function $R^{(L)}$ there by the value it takes as $u\to\infty$ along this line, and thus see that the remainder function asymptotes to this value at $u,v \to \infty$ more and more slowly as we go up in loop order. This behavior happens because the subleading power in $1/u$ is accompanied by up to $2L-1$ large logarithms at high loop orders.

\subsection{Special Points}
\label{sec:special_points}

We also study several points at which the form factor evaluates to interesting analytic constants, such as MZVs,
alternating sums, and cyclotomic zeta values~\cite{Ablinger:2011te}
including $6^{\rm th}$ roots of unity~\cite{HyperlogProcedures,Caron-Huot:2019bsq}. We do so partially in the $\mathcal{E}_c^{(L)}$ normalization introduced in section~\ref{sec:coaction_principle}, which features in the antipodal duality, and partially in terms of $\mathcal{E}$, which is an element of the smaller function space $\mathcal{C}$.

The first point we consider is the point on the $(u,u,1-2u)$ line corresponding to $u \to \infty$. In this limit, the form factor evaluates to MZVs. Through eight loops, it is given by the values
\begin{align}
\mathcal{E}_c^{(1)}(\infty, \infty, -\infty) &= 8 \zeta_2  \, ,  \label{eq:Ec_inf_1} \\
\mathcal{E}_c^{(2)}(\infty, \infty, -\infty) &=  31 \zeta_4  \, , \label{eq:Ec_inf_2} \\
\mathcal{E}_c^{(3)}(\infty, \infty, -\infty) &= - 145 \zeta_6  \, , \label{eq:Ec_inf_3} \\
\mathcal{E}_c^{(4)}(\infty, \infty, -\infty) &= \frac{11363}{4} \zeta_8 + 120 f_{5,3} \label{eq:Ec_inf_4} \, , \\
\mathcal{E}_c^{(5)}(\infty, \infty, -\infty) &= - \frac{257981}{5} \zeta_{10} - 240 \zeta_2 f_{5,3} - 1560 f_{5,5} - 2688 f_{7,3} \label{eq:Ec_inf_5} \, , \\
\mathcal{E}_c^{(6)}(\infty, \infty, -\infty) &=  \frac{33524700283}{33168} \zeta_{12} + 11160 \zeta_4 f_{5,3} + 48 \zeta_2 \big(95 f_{5,5} + 196 f_{7,3} \big) \nonumber \\ 
  &\qquad + 21120 f_{5,7} + 37296 f_{7,5} + 48528 f_{9,3} \label{eq:Ec_inf_6}  \, , \\
\mathcal{E}_c^{(7)}(\infty, \infty, -\infty) &= - \frac{1469870363}{80} \zeta_{14} - 302940 \zeta_6 f_{5,3} - 24 \zeta_4 \big(10265 f_{5,5} + 16912 f_{7,3} \big) \nonumber \\
  &\qquad - 96 \zeta_2 \big(1070 f_{5,7} + 1967 f_{7,5} + 2691 f_{9,3} \big) - 545664 f_{7,7}  \nonumber \\
  &\qquad - 319840 f_{5,9} - 722592 f_{9,5} - 1241760 f_{11,3} - 24960 f_{5,3,3,3} \label{eq:Ec_inf_7}  \, ,  \\
\mathcal{E}_c^{(8)}(\infty, \infty, -\infty) &= \frac{10561356154966997}{22917312} \zeta_{16} + 8548210 \zeta_8 f_{5,3} + 12 \zeta_6 \big(633995 f_{5,5} + 996436 f_{7,3} \big)  \nonumber \\ 
  &\qquad  + 48 \zeta_4 \big( 122940 f_{5,7} + 201929 f_{7,5} + 236967 f_{9,3} \big) \nonumber \\
  &\qquad  + 192 \zeta_2 \big( 22820 f_{7,7} + 13915 f_{5,9} + 29544 f_{9,5} + 57278 f_{11,3} + 1500 f_{5,3,3,3} \big) \nonumber \\
  &\qquad  + 9122624 f_{7,9} + 11543472 f_{9,7} + 5153280 f_{5,11} + 19603536 f_{11,5} \nonumber \\
  &\qquad   + 23915376 f_{13,3} + 371520 f_{5,3,3,5} + 400320 \big(f_{5,5,3,3} + f_{5,3,5,3} \big) \nonumber \\
  &\qquad  + 825216 f_{7,3,3,3}   \, , \label{eq:Ec_inf_8} 
\end{align}
where we have expressed everything in terms of an $f$-alphabet~\cite{Brown:2011ik}, which means that the coaction $\Delta$ acts by deconcatenation. 
A software implementation for working with these objects, which uses the same $f$-alphabet basis, is provided in ref.~\cite{HyperlogProcedures}. We have provided the value of the form factor at this point in terms of MZVs in the ancillary file {\tt AntipodePointsSummary} of ref.\ \cite{Dixon:2021tdw}.  This point is antipodally dual to the point $(\hat{u},\hat{v},\hat{w})=(1,1,1)$ for the six-point amplitude.  Notice that the coefficients are all integers, except for the coefficients of the pure even Riemann zeta values $\zeta_{2L}$.

\begin{table}
\begin{center}
\begin{tabular}{|l|c|c|c|c|}
\hline\hline 
\rule{0pt}{4ex}    
$L$ & $\frac{R^{(L)}(\frac{1}{3},\frac{1}{3},\frac{1}{3})}{R^{(L-1)}(\frac{1}{3},\frac{1}{3},\frac{1}{3})}$ & $\frac{{\mathcal{E}}^{(L)}(\frac{1}{3},\frac{1}{3},\frac{1}{3})}{{\mathcal{E}}^{(L-1)}(\frac{1}{3},\frac{1}{3},\frac{1}{3})}$ & $\frac{{\mathcal{E}_c}^{(L)}(\frac{1}{3},\frac{1}{3},\frac{1}{3})}{{\mathcal{E}_c}^{(L-1)}(\frac{1}{3},\frac{1}{3},\frac{1}{3})}$ & $\Gcusp^{(L)}/\Gcusp^{(L-1)}$ \\[.3cm]
\hline\hline
2  & --                      & $-$7.58283    & $-$6.13027 & $-$3.28987 \\
3  &  $+$56.20093  & $-$8.62198    & $-$8.79852 & $-$7.23771 \\
4  &  $-$19.93470   & $-$10.12258  & $-$11.12260 & $-$9.84237 \\
5  &  $-$15.97787   & $-$11.42376  & $-$12.46885 & $-$11.30950 \\
6  &  $-$15.20931   & $-$12.34388  & $-$13.21992 & $-$12.16022 \\
7  &  $-$15.00749   & $-$12.96338  & $-$13.68659 & $-$12.71784 \\
8  &  $-$14.96709   & $-$13.39441  & $-$14.00928 & $-$13.12172 \\
\hline\hline
\end{tabular}
\caption{\label{tab:R_symmax} The ratio of $R$, $\mathcal{E}$, and $\mathcal{E}_c$ at sequential loop orders, 
  evaluated at the symmetric point $(u,v,w)=(\frac{1}{3},\frac{1}{3},\frac{1}{3})$ in the Euclidean
  region. We also give the ratio of the cusp anomalous dimension at sequential loop orders for comparison.}
\end{center}
\end{table}

Other points where the form factor evaluates to nice constants include the point $(1,1,-1)$, where the form factor can be expressed in terms of alternating sums, and the points $(-1,-1,3)$ and ($\frac13, \frac13, \frac13$), where the form factor evaluates to cyclotomic zeta values \cite{Ablinger:2011te} including $6^{\rm th}$ roots of unity \cite{Caron-Huot:2019bsq}. The analytic expressions for these values are also provided in the file {\tt AntipodePointsSummary}. In Table~\ref{tab:R_symmax}, we show the numerical value of the remainder function, $\mathcal{E}$, and $\mathcal{E}_c$ at successive loop orders, evaluated at the point $(\frac13, \frac13, \frac13)$. We expect these ratios to asymptotically approach $-16$,
mainly because that is the inverse of the radius of convergence of the perturbative series for the cusp anomalous dimension $\Gcusp(g^2)$.  We see that $\mathcal{E}$ and $\mathcal{E}_c$ indeed appear to be approaching $-16$, even more quickly than the equivalent ratio for the cusp anomalous dimension.  The values for the remainder function are actually moving slightly further from $-16$; presumably that trend will change by nine loops.
  
The space of constants that $\mathcal{E}_c$ evaluates to at these special points matches the space of constants that the six-point amplitude evaluates to on the other side of the antipodal duality, as tabulated in Table I of ref.~\cite{Dixon:2021tdw}. Moreover, as observed there, the antipode acts on the elements of the $f$-alphabet basis used in eqs.~\eqref{eq:Ec_inf_1}--\eqref{eq:Ec_inf_8} by reversing the letters and multiplying each word of length $w$ by $(-1)^w$. Thus, up to factors proportional to $i \pi$ (on which the antipode is not defined), one can easily read off the value of the six-point amplitude from the expressions in the $f$-alphabet basis by applying the letter-reversal map.

\section{Conclusions}
\label{sec:conclusions}

In this paper, we have computed the three-point form factor of the chiral part of the stress tensor supermultiplet in planar $\mathcal{N}=4$ sYM theory at six, seven and eight loops. As far as we are aware, this is the highest perturbative order to which quantities depending on two generic variables are known in a four-dimensional unitary quantum field theory.  As such, our results provide high-order insights for a wealth of kinematics, some of which were explored in section~\ref{sec:special_kinematics}. For example, Figure~\ref{Fig:ELsym} shows a remarkable flattening of successive-loop ratios at high loop order, indicating that the finite radius of perturbative convergence of the cusp anomalous dimension also holds for generic scattering kinematics, at least when $u$ and $v$ are not too large.  Moreover, according to the principle of maximal transcendentality~\cite{Kotikov:2001sc,Kotikov:2002ab,Kotikov:2004er,Kotikov:2007cy}, our results are expected to provide the maximally transcendental (highest weight) piece of the remainder function for the corresponding Higgs-to-three-gluon amplitude in QCD in the limit of large top-quark mass.

To calculate this form factor, we have employed a bootstrap approach that closely follows the methods used to compute this quantity at lower loop orders~\cite{Brandhuber:2012vm,Dixon:2020bbt}. However, in the present paper we have further augmented these methods by observing and exploiting an extended set of ES-like relations~\eqref{eq:adjacency_restrictions_2}, a restriction on the triple sequences of letters at all depths in the symbol (given in eq.~\eqref{eq:triple_adjacency_restriction}), and an enlarged set of multiple-final-entry conditions. We have also used an improved normalization scheme in which the finite form factor belongs to a smaller space of polylogarithmic functions.

As part of our bootstrap procedure, we have used the FFOPE to compute the $T^2$ and $T^4$ contributions to the near-collinear limit of the form factor at six, seven, and eight loops. As summarized in Table~\ref{tab:Cparameters}, only some of the information contained in the $T^2$ term was needed to fix the last free coefficients in our ansatz. The fact that our form factors match the rest of the $T^2$ data, and all of the $T^4$ data, therefore provides a very non-trivial cross-check on our results. 

It is interesting to note that fewer coefficients needed to be fixed by FFOPE data at eight loops than at seven loops. This improved situation stems from the increasing power of the multiple-final-entry conditions. As recorded in the last line of Table~\ref{tab:finalentryconditions}, the number of new, independent multiple-final-entry conditions seems to grow (on average) with the number of final entries considered. Furthermore, the space of multiple final entries is observed to saturate at one higher weight at each perturbative order, and thus we get new constraints each time we compute one more loop.  In the present paper, we have used the $(L-1)$-final-entry conditions only after first observing them at $(L-1)$-loop order.

Just recently, we observed the remarkable fact that the three-point form factor studied in this paper is dual to the six-point MHV amplitude in planar $\mathcal{N}=4$ sYM theory~\cite{Dixon:2021tdw}. It is worth emphasizing that the computations carried out in the present paper involved no input from this duality; the only input to our bootstrap are the conditions described in Table~\ref{tab:Cparameters}. Rather, the results of these bootstrap computations were used in ref.~\cite{Dixon:2021tdw} to check the duality observed there through seven loops (as the six-point MHV amplitude is only known through seven loops).  On the other hand, the duality can now be exploited at eight loops, in order to provide the symbol of the eight-loop six-point MHV amplitude on the parity-preserving surface. Completing this symbol to the full eight-loop amplitude in the bulk will require considerably less linear algebra than computing it from scratch.

The antipodal duality between the three-point form factor and the six-point amplitude is particularly surprising, as the antipode reverses the order of the letters in the symbol and thus switches first and last entries, which have completely different physical interpretations. Currently, we do not see any physical reason why this duality should hold; it will be interesting to see whether it extends to further cases -- other theories, other operators, or different numbers of external states -- and whether it can be derived from first principles.

As discussed in section \ref{sec:duality}, the antipodal duality relates the multiple-final-entry conditions we have utilized to bootstrap the three-point form factor to the first-entry conditions and ES conditions of the six-point amplitude, which are known at all weights, as well as some additional constraints starting at weight eight  due to the knowledge of the OPE structure.%
\footnote{While it may be possible to prove the single-final-entry conditions using the $\bar{Q}$ equation for closed polygonal Wilson loops~\cite{Caron-Huot:2011dec}, it seems unlikely that this equation can also be used to understand the multiple-final-entry conditions.}
In future work, it would be interesting to exploit the duality to impose the part of the full $2L$-final-entry conditions that follow from the duality already at $L$-loop order.  It is tempting to believe that in this way the form factor at nine and higher loops is so constrained that little to no input from the near-collinear limit is required.   

It was also pointed out in section~\ref{sec:duality} that the antipodal duality relates the ES-like conditions~\eqref{eq:adjacency_restrictions_1} and~\eqref{eq:adjacency_restrictions_2} to the ES conditions of the six-point amplitude. 
In the absence of a derivation of this duality, it would be interesting to also understand the physical origin of the ES-like conditions more directly. A pair of intriguing observations on this topic were made in refs.~\cite{Dixon:2020bbt} and \cite{Chicherin:2020umh}, where the relations~\eqref{eq:adjacency_restrictions_1} were simultaneously first observed. In the former reference, it was shown that the class of pentabox ladder integrals $\Psi^{(L)}$ defined in ref.~\cite{Drummond:2010cz} naturally appear in the form factor space, and that the seven-particle ES conditions imply the form factor ES-like conditions~\eqref{eq:adjacency_restrictions_1}; here we observe that the ES-like conditions~\eqref{eq:adjacency_restrictions_2} are also implied by the seven-particle ES relations. See Appendix~\ref{sec:pbladders} for further details. Conversely, in ref.~\cite{Chicherin:2020umh}, it was shown that the symbol alphabet $\cL_u$ matches the cluster coordinates of the $C_2$ cluster algebra, and that a subset of the related cluster adjacency restrictions also provide the ES-like conditions~\eqref{eq:adjacency_restrictions_1}. However, still preferable to both of these observations would be an understanding of whether the ES-like conditions on $\mathcal{C}$ arise from physical principles that directly constrain the form factor. 

At higher points, the form factor for the chiral stress-tensor multiplet in planar $\mathcal{N}=4$ sYM theory is only known to one loop~\cite{Brandhuber:2010ad}. Thus, it would be especially interesting to extend the bootstrap methods used in this paper to higher particle multiplicity. While less is currently known about the spaces of functions that are relevant for these higher-point form factors, some of the integrals that contribute to the four-point form factor have recently been computed~\cite{Abreu:2020jxa,Syrrakos:2020kba,Canko:2020ylt,Abreu:2021smk,Chicherin:2021dyp,Kardos:2022tpo}. These results provide important information about the symbol letters that are expected to appear in four-point kinematics. 

Higher particle multiplicities are also the most direct direction for a potential extension of the antipodal duality between form factors and amplitudes.
The two-loop MHV amplitude is known for any number of particles~\cite{Caron-Huot:2011zgw,Golden:2014xqf,Golden:2021ggj} and is known to exhibit many interesting types of cluster-algebraic structure~\cite{Golden:2013xva,Golden:2014pua,Golden:2014xqa,Drummond:2017ssj,Golden:2018gtk,Golden:2019kks}. As a result, a great deal can be quickly learned about two-loop form factors if they can be related to amplitudes, even in restricted kinematics. Then one might be able to use the results for restricted kinematics as boundary conditions for determining the form factors in full kinematics. Importantly, the FFOPE also describes these higher-point form factors, and thus it can be used as an independent cross-check if and when conjectural results become available.


\acknowledgments We would like to thank Matt von Hippel, Andy Liu,
Oliver Schnetz, and Chi Zhang for useful discussions. This research
was supported by the US Department of Energy under contract
DE--AC02--76SF00515, and in part by the National Science Foundation
under Grant No.~NSF PHY--1748958.  AJM and MW were supported in part
by the ERC starting grant 757978 and grant 00015369 from Villum
Fonden. MW was additionally supported by grant 00025445 from the
Villum Fonden.  \"OG is supported by the UKRI/EPSRC Stephen Hawking
Fellowship EP/T016396/1. \"OG would like to thank the Isaac Newton
Institute for Mathematical Sciences for support and hospitality during
the Cluster Algebras and Representation Theory programme when work on
this paper was undertaken. This work was supported by EPSRC grant
number EP/R014604/1. LD thanks NYU, the Kavli Institute for
Theoretical Physics, and the DLITP for hospitality during this
work. The authors acknowledge the use of the IRIDIS High Performance
Computing Facility, and associated support services at the University
of Southampton, in the completion of this work. All data and the plots
in this paper follow from the results provided in the paper and in the
ancillary files. No other data has been used.
A figure was made with {\sc Jaxodraw}~\cite{Binosi:2008ig}.


\appendix

\section{Pentabox Ladders}
\label{sec:pbladders}

As observed in ref.~\cite{Dixon:2020bbt}, there exists an infinite sequence of dual-conformally invariant integrals that are expected to live in $\mathcal{C}$, namely the pentabox ladders $\Psi^{(L)}$.  These $L$-loop integrals consist of $L-2$ scalar boxes forming a ladder, capped off on one side by a pentagon integral with three external massless legs, and which is assigned a chiral loop-momentum-dependent numerator factor; the other end is capped by a scalar box with two external massive legs.  A differential equation has been derived that relates these integrals at adjacent loop orders~\cite{Drummond:2010cz}, which has been solved to all orders~\cite{Caron-Huot:2018dsv}.  The solution is phrased in terms of the double pentaladder integral $\Omega^{(L)}(\hat{u},\hat{v},\hat{w})$, which is related to the pentabox ladder by
\be 
 \Omega^{(L)}(\hat{u},\hat{v},\hat{w})\Bigl|_{\hat{w}\to0}\ =\ \Psi^{(L)}(\hat{u},\hat{v}).
\label{PsifromOmega}
\ee
The pentabox ladders live in the space of (extended Steinmann-satisfying)
heptagon functions for seven-point amplitudes~\cite{Dixon:2016nkn},
because the two massive legs of the box can be split into four massless legs,
plus the three massless legs attached to the pentagon.
However, the pentaladders should also live in a suitable space of form factor
functions, because taking the dual point
between the two massive legs to infinity maps the two relevant dual conformal cross ratios $\hat{u}$ and $\hat{v}$ to the form factor variables $u$ and $v$; see Appendix A of ref.~\cite{Dixon:2020bbt} for more details.

Interpreted as functions of the form factor variables, the pentabox ladders only involve five of the six letters in the alphabet $\mathcal{L}_a$, namely
\be \label{eq:pbletters}
\cL_{\Psi}\ =\ \{ a,b,c,d,e \}  \, .
\ee
They obey the same final-entry conditions as the three-point form factor,
\be \label{eq:pbfe}
\Psi^a\ =\ \Psi^b\ =\ \Psi^c\ =\ 0,
\ee
and thus they have only two independent final entries, $\Psi^d$ and $\Psi^{e}$.
They also satisfy the double-final-entry relations 
\bea
\Psi^{a,d} &=& \Psi^{e,d} = \Psi^{b,e} = \Psi^{d,e} = \Psi^{b,d} = \Psi^{a,e} = 0,
\nonumber\\
\Psi^{c,e} &=& \Psi^{c,d} = \Psi^{(L-1)} \,,
\label{eq:pbdfe}
\eea
which leave three independent double-final entries, $\Psi^{d,d}$, $\Psi^{e,e}$,
$\Psi^{c,d}$.  The last of these double final entries is equal to the $(L-1)$-loop pentaladder integral,
and in fact \eqns{eq:pbfe}{eq:pbdfe} are equivalent to the second-order
differential equation satisfied by this class of integrals~\cite{Drummond:2010cz}.

Starting from the known all-orders solution for $\Omega^{(L)}(\hat{u},\hat{v},\hat{w})$
and taking the limit~\eqref{PsifromOmega}, we have confirmed that the functions $\Psi^{(L)}$ exist within the space $\mathcal{C}$ 
through eight loops. Thus, the pentaladder integrals obey the same 
ES-like relations~\eqref{eq:adjacency_restrictions_1} and~\eqref{eq:adjacency_restrictions_2}, constraints on adjacent triples~\eqref{eq:triple_adjacency_restriction}, and coaction principle as the form factor $\mathcal{E}^{(L)}$. Note that, while the space of functions defined by $\Psi$ and all of its coproducts is contained in $\mathcal{C}$, it is also much smaller, as these coproducts span a space that has a dimension of just $w+1$ at weight $w$~\cite{Caron-Huot:2018dsv}.  One can also consider the larger space corresponding to all ES-satisfying heptagon functions that can be constructed out of the alphabet $\mathcal{L}_\Psi$; this space also turns out to be contained in $\mathcal{C}$ at symbol level. 

Interestingly, although the double pentaladder integrals $\Omega^{(L)}(\hat{u},\hat{v},\hat{w})$ live in the hexagon function space, and contribute to the six-point MHV amplitude, the antipodal duality does \emph{not} map them into $\mathcal{C}$.  The pair and triple relations are all satisfied, thanks to \eqns{eq:mapadj_1}{eq:mapadj_2}.  However, the multiple-final-entry conditions for $\Omega^{(L)}(\hat{u},\hat{v},\hat{w})$ do not precisely correspond to those of the six-point MHV amplitude.  Hence, after applying the duality, the initial (multiple) entries of the integral are not in $\mathcal{C}$, at least for the four loop orders we have inspected.  More precisely, the dual of $\Omega^{(1)}$ does not satisfy the first entry condition; the dual of $\Omega^{(2)}$ satisfies that condition, but not the first two entry conditions of $\mathcal{C}$; the dual of $\Omega^{(3)}$ satisfies the first two, but not the third; and the dual of $\Omega^{(4)}$ satisfies the first three, but not the fourth.  The lesson seems to be that the full amplitude behaves better than individual integrals under the duality.


\providecommand{\href}[2]{#2}\begingroup\raggedright\endgroup

\end{fmffile}

\end{document}